\documentclass[prl,aps,reprint,showpacs,showkeys,floats,floatfix,superscriptaddress,longbibliography]{revtex4-2}

\usepackage[english]{babel}
\usepackage{color}
\usepackage{graphicx}
\usepackage[caption=false]{subfig} 
\usepackage{amsmath,amssymb,bm}
\usepackage[version=3]{mhchem}
\usepackage{verbatim}
\usepackage{multirow}
\usepackage{dcolumn}
\usepackage{float}
\usepackage{nicefrac}
\usepackage{siunitx}
\DeclareSIUnit[number-unit-product = {\,}]{\au}{a.u.}
\DeclareSIUnit[number-unit-product = {\,}]{\kJmol}{\kilo\joule\per\mol}
\usepackage[colorlinks,allcolors=black,citecolor=blue,urlcolor=blue]{hyperref}
\usepackage{booktabs}

\newcommand{\pmeth}{CH$_5^+$}
\newcommand{\meth}{CH$_4$}
\newcommand{\he}{$^4$He}

\newcommand{\ccdots}{\makebox[1em][c]{$\cdot$\hfill$\cdot$\hfill$\cdot$}}

\begin{document}
\title{Manifestations of Local Supersolidity of $^\text{4}$He around a Charged Molecular Impurity}
\author{Fabien Brieuc}
\email{fabien.brieuc@cea.fr}
\affiliation{Lehrstuhl f\"ur Theoretische Chemie,
 Ruhr-Universit\"at Bochum, 44780 Bochum, Germany}
\affiliation{Present address: Laboratoire Mati\`ere en Conditions Extr\^emes,
Universit\'e Paris-Saclay, CEA, 91680 Bruy\`eres-le-Ch\^atel; 
CEA, DAM, DIF, F-91297 Arpajon, France}
\author{Christoph Schran}
\email{christoph.schran@rub.de}
\affiliation{Lehrstuhl f\"ur Theoretische Chemie,
 Ruhr-Universit\"at Bochum, 44780 Bochum, Germany}
\affiliation{Present address: Yusuf Hamied Department of Chemistry,
University of Cambridge, Lensfield Road, Cambridge CB2 1EW, UK}
\author{Dominik Marx}
\email{dominik.marx@rub.de}
\affiliation{Lehrstuhl f\"ur Theoretische Chemie,
 Ruhr-Universit\"at Bochum, 44780 Bochum, Germany}
\date{\today}

\begin{abstract}
A 
frozen,
solid helium core,
dubbed snowball,
is 
typically observed around 
cations
in liquid helium.
Here we 
discover, 
using 
path integral simulations, that 
around a cationic molecular impurity, 
protonated methane,
the \he{} atoms are
indeed strongly localized 
akin to snowballs 
but 
still participate in vivid bosonic exchange 
induced by the
ro-vibrational motion of 
the impurity.
Such
combination of solid-like order with 
pronounced
superfluid response in the first helium shell 
indicates that 
manifestations of
local 
supersolid behavior 
of \he{} 
can be 
induced
by
charged molecules.
\end{abstract}

\maketitle

\section{Introduction}
Since the ground-breaking discovery of superfluidity in liquid 
\he{}, generations 
of scientists have raised the question whether such superfluid 
response can persist in the solid phase, i.e. 
\textit{{Can a {Solid} {Be} ``{Superfluid}''?}}~\cite{leggett_can_1970}.
Such a counter intuitive state, combining solid-like order with a 
finite
superfluid fraction~\cite{boninsegni_colloquium_2012}, has been theoretically
proposed 
as early as 
1969~\cite{%
andreev_quantum_1969,
reatto_bose-einstein_1969%
}.
More recently, it has been suggested experimentally that supersolid 
states can appear in model systems, such as Bose-Einstein condensates of 
atomic gases at ultralow~temperatures~\cite{li_stripe_2017,ferlaino_supergas_2021,bland_supergas_2022}.
Computationally, evidence of a supersolid phase has been reported for atomic deuterium~--
yet 
at ultrahigh pressure conditions~\cite{parrinello_superD_2022}. 
When it comes to finite systems
substantial understanding was provided by theoretical work
leading to novel experiments.
In fact, pioneering
path integral simulations have revealed that small para-H$_2$
clusters of specific
size can exhibit
spatial
localization of 
these bosonic species at sufficiently low temperatures, 
reminiscent of 
a solid,
combined with some remaining 
bosonic exchange~\cite{sindzingre_superfluidity_1991,mezzacapo_possible_2011},
thus suggesting 
supersolid behavior.
But different from para-H$_2$,
such phenomena cannot appear in pure \he{} clusters
since they remain liquid even in the ground state
given that the very weak He$\cdots$He  interactions  
are a factor of 3~smaller. 
Overall, the 
existence of 
supersolid properties 
in the specific case of \he{}--systems,
both extended and finite, remains
controversial
\cite{kim_probable_2004,%
balibar_enigma_2010,
kim_absence_2012,%
boninsegni_colloquium_2012,
kim_2dhe_2021}.
Here,
we answer the question whether 
\textit{manifestations of supersolid behavior} 
of \he{} can be found 
\textit{locally} 
in the first shell
around charged molecular impurities
in 
finite
\he{}--systems 
--- a situation that would
potentially
enable 
experimental
verification 
akin to discovering \textit{``microscopic manifestation of superfluidity''} 
around neutral dopant molecules~
\cite{grebenev_superfluidity_1998,toennies_spectroscopy_1998,toennies_superfluid_2004}.
Indeed, 
such
impurities 
in liquid helium have been shown to be powerful probes of 
\textit{``manifestations of superfluid behavior''},
a concept introduced by seminal theory work~\cite{sindzingre_path-integral_1989},
and validated experimentally a decade later~\cite{%
grebenev_superfluidity_1998,%
toennies_spectroscopy_1998,%
toennies_superfluid_2004%
}.
Charged impurities
usually interact strongly with helium, 
creating 
a frozen, solid-like core around the impurity, 
referred to as a 
snowball~\cite{atkins_ions_1959}.
This snowball effect has been extensively studied both 
theoretically~\cite{%
nakayama_theoretical_2000,%
buzzacchi_alkali_2001,%
galli_path_2011%
} and experimentally~\cite{poitrenaud_precise_1972,%
muller_alkali-helium_2009,bartl_size_2014} 
in the case of atomic 
cationic impurities,
and has been linked to a 
local {\em disappearance of superfluidity} within the 
frozen
\he{} atoms of the snowball due to their strong 
spatial
localization~\cite{%
bartis_mobility_1977,nakayama_theoretical_2000,galli_path_2011}.
Molecular impurities have 
also
been immersed 
particularly
in helium nanodroplets and smaller 
\he{}
clusters~\cite{toennies_superfluid_2004}.
Their rotational excitations have been used 
to probe the 
local
microscopic superfluid response 
showing that 
\textit{``manifestations of superfluid behavior''}
can 
indeed
be found in finite systems as small as 
\mbox{about ten \he{}}
atoms only~\cite{%
tang_quantum_2002,%
xu_spectroscopic_2003,%
mckellar_spectroscopic_2006%
}.
In addition,
quantum simulations have been pivotal
in elucidating the impact of neutral molecular impurities
on the helium environment
~\cite{%
kwon_atomic-scale_1999,%
draeger_superfluidity_2003,%
moroni_structure_2003,%
tang_bridging_2004,%
paesani_onset_2005%
}.
Their
impact 
is usually smaller than 
ions due to 
weaker interactions with the solvent.
However, around some of the most strongly interacting neutral 
molecules, such as \ce{SF6}, the first 
solvation shell is composed of more 
localized \he{} atoms which is linked to a \textit{reduction} of the 
superfluid fraction in a way that is reminiscent of the snowball 
effect around ions~\cite{%
kwon_atomic-scale_1999,%
duminuco_local_2000}.
Yet, it has been 
shown that rigid-body rotation 
can lead to a non-negligible \textit{enhancement} of the superfluid response 
in the first shell~\cite{%
kwon_quantum_2000,%
blinov_path_2004,%
zillich_path_2005,%
markovskiy_path_2009}.
Pioneering work on the superfuid response of \he{} around 
C$_{20}$ has even hinted at 
something like
nanoscale supersolidity but, 
akin to \mbox{para-H$_2$} clusters, 
only for specific 
(``magic'')
helium numbers
between 28 and 31 atoms,
whereas the phenomenon quickly vanishes 
when adding 
more \he{} atoms~\cite{kwon_superfluidity_2010}.
Similar effects have also been seen around 
the completion of the second layer of
helium adsorbed on graphite~\cite{crowell_superfluidity_1996,%
nyeki_supersolid_2017,
nyeki_intertwined_2017}.
In 
that
long-standing
quest of finding manifestations of supersolid behavior
in \he{}, 
this leads to the question 
whether fast rotation of 
a strongly interacting 
molecular
impurities could induce a 
pronounced 
superfluid response in a 
frozen,
solid-like helium shell 
and, thus, could be used as seeds for supersolid behavior in the bulk.
To answer this 
fundamental
question, we set out to study helium solvation of
an ionic molecular impurity, 
protonated methane,
using 
quantum simulations.
Why \pmeth{}~?
Firstly, 
its interaction with \he{} is about 
four times 
stronger than
other strongly interacting 
neutral
species. 
Secondly,
\pmeth{} is a prototype of 
the class of so-called 
fluxional molecules,
being
subject to large-amplitude motion leading to a full delocalization 
of 
its hydrogens due to pseudo-rotational motion~\cite{marx_structural_1995}. 
The combination of these 
intramolecular 
pseudo-rotations with the standard 
rotations of the molecule leads to 
entangled 
\mbox{SO(5)} ``superrotational motion''~\cite{schmiedt_collective_2016}. 
This complex and rich 
``hydrogen scrambling''
dynamics remains
unperturbed under helium solvation~\cite{davies_infrared_2021} 
in agreement with our present findings.
Thirdly,
an intricate coupling has recently been discovered 
between 
the complex ro-vibrational 
motion of 
\pmeth{}
and bosonic exchange
in the \textit{microsolvation limit} with up to only
four \he{} atoms~\cite{uhl_quantum_2019}.

We 
study 
quantum solvation 
and superfluid response 
of \pmeth{} in 
\he{} nano\-clusters of up to 60~helium atoms
and unveil a novel phenomenon
that is unknown in the microsolvation limit. 
These 
results 
are referenced to  \meth{} in helium~\cite{markovskiy_path_2009} 
to compare to this ``ordinary cousin'' of \pmeth{}.
Contrary to \pmeth{}, \meth{} is 
a standard quasi-rigid molecule subject to
small-amplitude motion that is well described by 
quasi-harmonic deviations from a unique equilibrium structure.
Moreover, 
as most neutral species, 
\meth{} does not feature such 
strong interactions as \pmeth{} with helium.
\section{Methods and computational details}
All simulations of \pmeth{} and \meth{} solvated in clusters composed
of $n=1$ up to 60 \he{} atoms have been carried out at $T=0.5$\,K using  
finite temperature bosonic path integral techniques.
This approach takes into account 
the full molecular flexibility at essentially converged coupled cluster 
level as recently reviewed in Ref.~\citenum{brieuc_converged_2020}. 
Accordingly, we used a hybrid \mbox{PIMD/ PIMC} approach~\cite{%
walewski_reactive_2014} in which helium is sampled using 
path integral Monte Carlo (PIMC)~\cite{ceperley_path_1995}
to account for the bosonic nature of \he{}, whereas 
the molecule is described using path integral molecular dynamics (PIMD).
More precisely, the configuration and permutation space of bosonic 
helium is sampled using the continuous-space worm algorithm~\cite{%
boninsegni_worm_2006,boninsegni_worm_2006_2} (in the specific canonical
variant introduced in the appendix of Ref.~\citenum{brieuc_converged_2020}), 
while the configuration 
space of the impurity is sampled using the path integral quantum 
thermal bath (PIQTB) technique~\cite{brieuc_quantum_2016}
as adapted and validated for path integral simulations 
at very low temperatures~\cite{schran_zundel_2018}.
Path integral convergence is achieved by describing
the helium density matrix within the pair density approximation~\cite{ceperley_path_1995}
using a high-temperature matrix computed at $T=80$\,K resulting in a path 
integral discretization of 160~beads at $0.5$\,K, while the path integral
is discretized using 640~beads for the molecule in conjunction with the
PIQTB thermostat.
The 
results
reported here have been obtained by averaging over 
20~independent runs propagated using a formal PIMD time step of 0.25\,fs for 
$4\cdot 10^5$~steps corresponding to a trajectory length of 
formally 100\,ps for each of the independent runs.
Between two PIMD steps, helium was sampled using at least $1 \cdot 10^{5}$ 
PIMC moves per helium atom.
All 
interactions involving \pmeth{} and \meth{}
are represented using 
highly
accurate
neural network potentials 
trained to essentially converged coupled cluster 
electronic structure calculations 
\cite{schran_high-dimensional_2017,schran_automated_2020}.
We refer to the SM
Sec.~I 
for more information on the methodology of these neural network 
potentials as well as for comprehensive benchmarks.
To study the superfluid response of \he{} around the molecule, we compute the
superfluid fraction $f_s$ of helium which quantifies the fraction of helium being in 
the superfluid state.
Within the two-fluid model of superfluidity, the total helium density is divided into a 
superfluid density $\rho_{\rm s}$ and a normal density $\rho_{\rm n}$ and the superfluid 
fraction is then obtained as the ratio $f_s=\rho_{\rm s}/\rho_{\rm n}$.
Various estimators have been developed to compute the superfluid fraction 
in path integral simulations. 
Here we use the so-called ``area estimator'' that has been developed specifically for 
finite size clusters~\cite{sindzingre_path-integral_1989} and which is based on the 
vectorial area of the exchange path, see Sec.~III.A of the SM for more detailed information.
One should note that this estimator is formally valid in the thermodynamic limit 
and in practice presents some limitations when dealing with very small clusters.
In particular, it gives a non zero superfluid fraction for a single helium atom.
This drawback can be corrected using a rescaled estimator, the so-called 
``exchange estimator''~\cite{mckellar_spectroscopic_2006}. 
We carefully tested the validity of the area estimator in our case, see Sec.~III.A and, 
Fig.~S14
of the SM, by comparing to the exchange estimator.

Moreover, in order to get some local information about the superfluid response of helium, 
we estimate the local superfluid density $\rho_{\rm s}(\vec{r})$ in two different ways.
The first estimator we use is based on the length of the exchange path.
Indeed, it is well-known that superfluidity is related to the presence of long exchange 
path~\cite{krauth_quantum_1996}.
It is thus possible to estimate $\rho_{\rm s}(r)$ by computing the density of long exchange 
paths~\cite{kwon_atomic-scale_1999}, $\rho_{\rm s}(\vec{r})=\sum_{p>l}^{N_{\rm He}} \rho_p(\vec{r})$, 
where the sum runs over
exchange paths exceeding a user-defined length of $l$.
In this work, we chose $l=6$,
i.e. all exchange paths involving more than 6 atoms are considered
to 
contribute
to the superfluid density.
We checked that the obtained superfluid densities remain qualitatively similar for different 
values of the cutoff length, see Sec.~III.B and, in particular 
Fig.~S16
of the SM.
The second estimator we use to compute superfluid densities is a generalization of the 
area estimator of the superfluid fraction, see Sec.~III.B of the SM for more details
including the expression and its derivation.

\section{Structural Properties}
\subsection{Molecular Structure}
\begin{figure}
   \centering
   \includegraphics{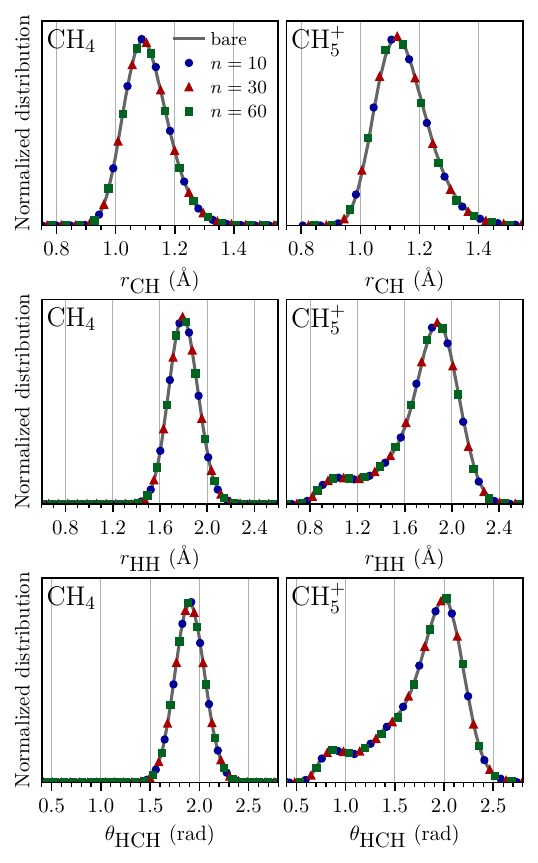}
   \caption{Molecular structure of \meth{} (left column) 
            and \pmeth{} (right column) solvated by different numbers of 
            helium atoms $n$ (symbols) and compared to the isolated bare 
            $n=0$ case (gray line) at $T=0.5$\,K. The distributions of 
            CH~distances, HH~distances and HCH~angles are presented 
            in the top, middle, and bottom panels, respectively.}
   \label{fig:mol-struct}
\end{figure}

The first question that arises 
is the impact of the solvent on the molecular 
properties, in particular their structure and dynamics.
It is usually assumed that the impact of helium is negligible,
due to the weak nature of its interaction with the molecule, 
and helium is thus considered one of the best solvents 
in particular for spectroscopic studies~\cite{toennies_spectroscopy_1998}.
The question of the impact of helium on fluxional molecules,
which exhibit fragile large-amplitude motion, has 
been recently studied for \pmeth{} microsolvated with up to 
only~four \he{} atoms~\cite{uhl_helium_2018}.
This study has revealed that the impact of helium on the 
molecular structure is indeed negligible even for the utmost
fluxional \pmeth{} molecule, and that helium does not seem to impact
the fluxionality and large-amplitude motion.
As clearly seen in the identical distribution functions
for key structural properties of \meth{} and \pmeth{}
in Fig.~\ref{fig:mol-struct}, the
impact of helium remains negligible, even if the
first helium solvation shell is fully closed with 16 
helium atoms,
probed here with $n=30$ \he{}. 
Moreover, no significant change in the molecular structure is 
observed when increasing the number of helium atoms 
even up to 60 \he{} around the molecule, in which case the 
molecule is entirely immersed in helium.
Thus, the findings of Ref.~\citenum{uhl_helium_2018} are not only 
valid in the extreme microsolvation limit but extend further to the 
fully solvated regime.
\subsection{Helium Solvation Structure}
\begin{figure*}
   \centering{}
\includegraphics{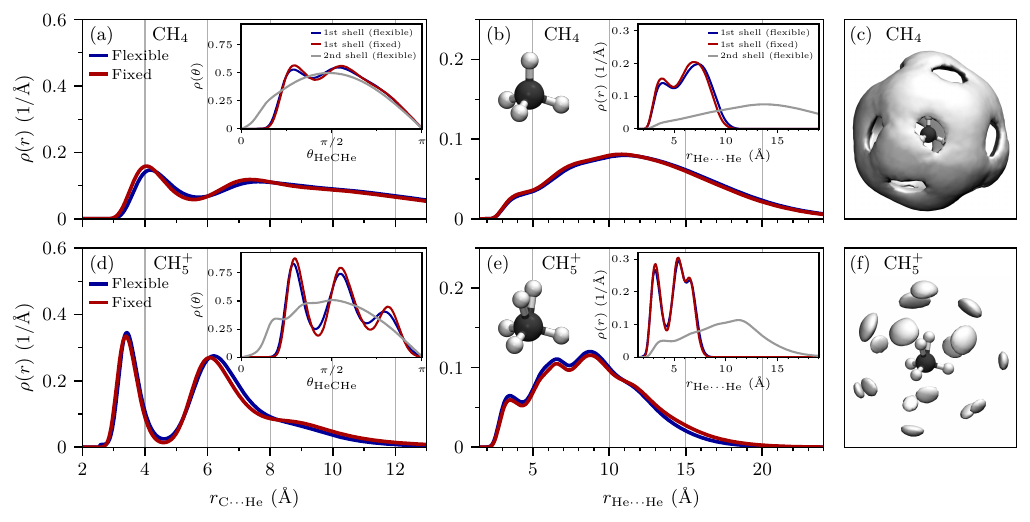}
   \caption{%
Distance
distribution functions of C{\ccdots}He 
            ((a) and (d)) and He{\ccdots}He ((b) and (e)) distances 
            as well as spatial distribution functions (SDFs) of \he{}
around the impurities
((c) and (f)) 
for
            \meth{}$\cdot$\he$_{60}$ 
(top row) 
            and \pmeth{}$\cdot$\he$_{60}$ 
(bottom row); 
data for the fully flexible and completely fixed impurities 
(with all constituting
nuclei frozen as point-particles at their equilibrium structures 
as depicted in the insets of~(b) and~(e))
are shown in blue and red, respectively.
The insets of~(a) and~(d) depict the angular 
            distribution functions of \he{} w.r.t. the carbon atom in the first 
            and second (gray) shell, whereas the insets of~(b) and~(e) show 
            the He{\ccdots}He distance distributions split into first and 
            second (gray) shell. 
The SDFs have been
computed w.r.t. fixed impurities using isovalues
corresponding to 
            a third of the respective maximum value 
which highlights the first \he{} shell. 
            }
   \label{fig:distrib}
\end{figure*}
The impact of neutral and charged impurities on the
surrounding helium is illustrated by the distributions
reported in Fig.~\ref{fig:distrib} for \meth{} versus \pmeth{}
solvated by 60 \he{} atoms.
The pronounced \pmeth{\ccdots}\he{} interaction
leads to 
large density modulations 
around the impurity 
and almost no 
interchange
of helium atoms
between the 
first and second solvation shell,
as seen by the density close to zero
in between the two 
peaks
in Fig.~\ref{fig:distrib}(d).
In contrast, the presence of \meth{} leads to 
mild 
modulations
even though 
a weakly defined, faint 
first shell can 
be identified
followed by a 
region of significant 
helium density 
allowing for easy
interchange of $^4$He between the first shell and beyond. 
These are the signatures 
of the much weaker interaction of $^4$He with CH$_4$ compared to CH$_5^+$
as a result of CH$_5^+$ being a charged molecule.
This significant impact of \pmeth{} on the solvent structure can also be seen
in the He{\ccdots}He distributions that 
indicates
a highly 
structured helium environment 
only around \pmeth{} whereas not much such structure is seen for CH$_4$. 
This is 
most pronounced
in 
the first shell, 
see inset of Fig.~\ref{fig:distrib}(e),
where the high density of 
about 0.11~\AA{}$^{-3}$,
which largely exceeds the freezing density of bulk helium,
results in
a solid-like order~---
the \he{} snowball as known from 
simple
monatomic cations. 
This solid-like structure of the first shell is confirmed by the 
angular distribution of helium around the carbon atom of \pmeth{},
shown in the inset of Fig.~\ref{fig:distrib}(d), that exhibits sharp 
peaks in the first shell.
Finally, the spatial distribution functions (SDFs) depicted in the 
rightmost column of Fig.~\ref{fig:distrib} clearly summarize 
what is observed based on these radial and angular distribution functions. 
The 
real-space structuring of the probability distribution of $^4$He
atoms around CH$_5^+$ versus CH$_4$ appears clearly: $^4$He atoms are 
significantly more localized around CH$_5^+$~(f) compared to CH$_4$~(c)
which indicates \emph{pronounced local translational and orientational order
of helium close to CH$_5^+$}.
This is what we call ``solid-like order'' in such a finite cluster
obviously without implying long-range periodicity as in an extended crystal.

Overall, this
analysis of the 
helium 
structure reveals 
the presence of a frozen 
first shell 
around \pmeth{}.
In contrast, this phenomenon is not present for \meth{}, in line with 
previous work~\cite{markovskiy_path_2009}, 
as supported by all data in panels~(a) to~(c) of Fig.~\ref{fig:distrib}.
Interestingly, the 
ro-vibrational
motion of the impurities,
even of the 
strongly interacting \pmeth{},
has negligible impact on the solvation shell structure
for sufficiently large clusters 
(compare blue to red lines in Fig.~\ref{fig:distrib}).
This behavior is 
different 
in small clusters 
(see SM~Fig.~S10),
for which the molecular motion has a significant impact
on the helium solvation structure.
A pictorial summary of the 
structural differences
is provided by the 
different
SDFs
in Fig.~\ref{fig:distrib}:
While the 3D~distribution of \he{} around \meth{} in panel~(c)
is 
broad
and smeared out
(``liquid-like''),
\he{} 
atoms are
localized 
at well-defined position 
(``solid-like'')
around
\pmeth{} in~(f).

\section{Superfluid properties}
\begin{figure}
   \centering{}
   \includegraphics{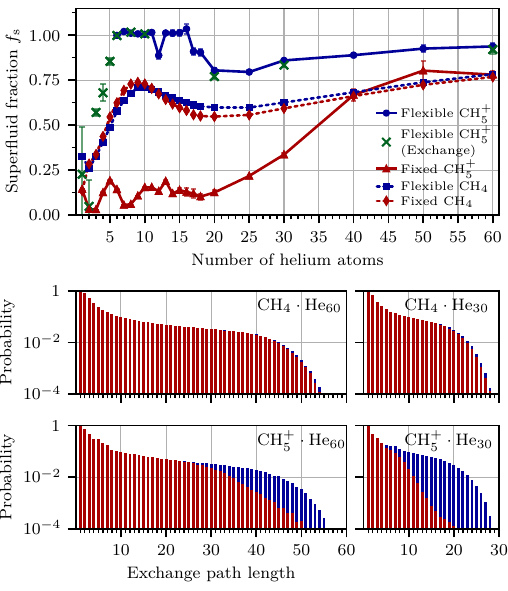}
   \caption{Top: Superfluid fraction $f_{\rm s}$ for \meth{}$\cdot$\he$_n$ 
            (dashed lines) and \pmeth{}$\cdot$He$_n$ (solid lines) 
as a function of $n$;
the ``exchange estimator'' has been used to compute the green crosses 
whereas all other data have been obtained from the ``area estimator''
as explained in SM~Sec.~III.A and validated in 
SM~Fig.~S14.
Bottom: Probability of finding at 
            least one exchange path of specific length
for $n=60$ (left) and $n=30$ (right) \he{} atoms.
Data for flexible and fixed impurities 
(see caption of Fig.~\ref{fig:distrib})
are shown in blue and red, respectively;
note that blue is superimposed by red where not visible. 
}
   \label{fig:sff_perm}
\end{figure}

How are these structural differences linked to
the superfluid response
of \he{},
$f_{\rm s}$,
as a function of the number $n$ of \he{} atoms?
Within 
the two-fluid model of superfluidity, 
the superfluid fraction
$f_{\rm s}$
is defined as the ratio between the superfluid density
$\rho_{\rm s}$ and the total density $\rho$.
It can be obtained directly from our quantum simulations
and, in addition, spatially decomposed 
in terms of local shell contributions based on the radial superfluid density $\rho_{\rm s}(r)$,
see SM~Sec.~III for methods.
Figure~\ref{fig:sff_perm} shows the global superfluid fraction
of the helium around \meth{} and \pmeth{} as a function of 
the number of helium atoms.
For \meth{}, 
$f_{\rm s}$ 
increases with $n$ 
before reaching a maximum at $n=9$ 
which corresponds to the maximum size for which all
helium atoms mostly belong to the
first shell, 
as shown in 
SM~Fig.~S12. 
After that, 
$f_{\rm s}$ drops initially 
due to the low density of additional helium
atoms 
outside the first shell
which hinders bosonic exchange.
Upon 
filling the second shell, 
$f_{\rm s}$
increases again, 
reaching about 0.78 for the largest cluster.
As for the helium densities,
the
ro-vibrational motion of \meth{}
has negligible impact on
$f_{\rm s}$
around this quasi-rigid and weakly interacting molecule
as fixing its nuclei (red 
diamonds compared to blue squares)
does not change $f_{\rm s}$
We note that we obtain very similar values of
$f_{\rm s}$ at 
$T=0.3$\,K (flexible: 0.91, fixed: 0.88) as reported for
12 helium 
atoms
around \meth{} in Ref.~\citenum{markovskiy_path_2009}.
The superfluid response 
around \pmeth{}
is distinctly different.
Despite the strongly localized helium 
in 
the first shell, 
recall Fig.~\ref{fig:distrib}(f) versus (c), 
a large superfluid fraction,
exceeding even the largest 
value obtained for the 
weakly interacting \meth{},
is found around
the \pmeth{} impurity. 
Moreover,
neglecting the 
ro-vibrational motion 
by fixing \pmeth{}
now greatly suppresses $f_{\rm s}$,
indicating almost no superfluid response
for clusters of size lower than around 30 \he{} atoms
(red triangles compared to blue circles) 
as expected from the traditional snowball picture.
In the limit of large clusters, the 
\emph{global}
superfluid fraction tends to 
unity,
which is expected regardless of the impurity since \he{} is indeed 
superfluid at this temperature and we thus retrieve the bulk limit.
In other words, for 
CH$_4$, the superfluid fraction does not change
whether we fix the molecule in its equilibrium structure or not.
This indicates that there is no impact of the ro-vibrational motion of CH$_4$
on the superfluid response of the surrounding helium.
In stark contrast, in the case of CH$_5^+$, the superfluid fraction values obtained
when fixing the molecule in space are extremely small compared to the large values found
when allowing for full flexibility of the molecule. 
This shows that, 
in the case
of protonated methane, the ro-vibrational motion of the molecule plays a
crucial role and actually considerably enhances the superfluid response of
the surrounding helium~--
despite the pronounced \emph{local translational and orientational order of helium 
observed in the first shell around CH$_5^+$.} 

\begin{figure*}
   \centering{}
   \includegraphics{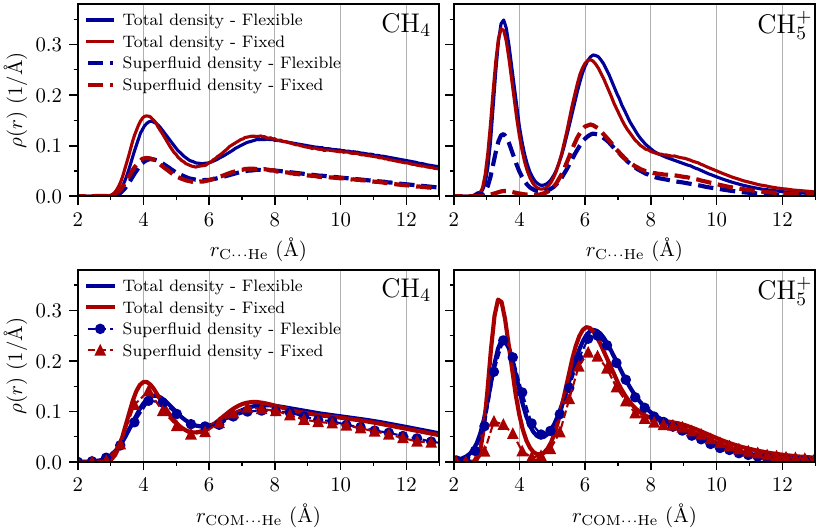}
\caption{
Local superfluid density $\rho_{\rm s}(r)$ computed using the 
``local exchange path estimator'' (top row -- dashed line) considering 
seven or more \he{} atoms ($l=6$), and using the ``local area path estimator'' 
(bottom row -- dashed line with symbols), together with the radial total 
density $\rho(r)$ (full line). Data for the flexible and fixed impurities 
(see caption of Fig.~\ref{fig:distrib}) are shown in blue and red, respectively.
Note that the definition of the radial density in the 
top and bottom panels is
slightly different: 
in the top panels
the distance $r$ is
defined with respect to the
central Carbon atom of the molecule, while in the
bottom panel it is with respect
to the center of mass of the molecule as indicated in the 
axis labels.
} \label{fig:sfd}
\end{figure*}

A deeper understanding 
can be obtained by studying 
the bosonic exchange path statistics
for flexible versus fixed molecular impurities, 
see Fig.~\ref{fig:sff_perm}.
The
ro-vibrational motion
of 
the fluxional 
\pmeth{} significantly 
enhances bosonic exchange and facilitates long exchange cycles that 
are known to be related to superfluidity~\cite{%
krauth_quantum_1996,%
kwon_atomic-scale_1999}.
No such impact
on bosonic exchange 
is observed for the
neutral 
\meth{}.
Studying the evolution of the superfluid fraction with cluster size 
$n \geq 6$ 
in Fig.~\ref{fig:sff_perm}, 
one can see that in the case of flexible \pmeth{}, 
$f_{\rm s} \approx 1$ for $n=6$ to $16$ which corresponds to the first 
shell being completely filled
(except for a drop at $n=12$ which is due to a
known topological phenomenon
as explained in the~SM based on 
Fig.~S15).
After that, the value of $f_{\rm s}$ is slightly reduced due to the
buildup of a second shell with a locally reduced \he{} density that 
disfavors 
bosonic exchange. 
Upon increasing $n$, the second shell gets 
filled and 
$f_{\rm s}$
increases accordingly, 
reaching  $f_{\rm s} \approx 0.9$ for $n=60$, 
see 
SM~Fig.~S12
for shell filling analysis.
Our analyses 
indicate
that the
first solvation 
shell of \he{} around flexible \pmeth{}
features maximum superfluidity as quantified by $f_{\rm s}\approx 1$.
Such 
pronounced 
superfluid response is found
despite the 
first shell being solid-like. 
Spatially-resolved insights into
the superfluid response of helium 
can be obtained
by defining a 
local superfluid density $\rho_{\rm s}(r)$ 
as presented in Fig.~\ref{fig:sfd}
for two different local estimators;
see SM~Sec.~III.B
for an extended discussion and validation.
Similarly to what was observed for the superfluid fraction,
the 
ro-vibrational motion of \meth{}
has negligible impact on the superfluid density from
the first shell up to the largest \mbox{C$\cdots$He} distances.
For \pmeth{}, a very
different
scenario is found:
Essentially zero
superfluid density is present in the
first \he{} shell around the fixed \pmeth{} molecule
compared to a significant superfluid response in the second shell
as clearly seen in the shell-resolved 
integrated 
superfluid densities in
Table~\ref{tab:sf-shell}.
This is a direct consequence of the solid-like order of the helium in 
the first shell~-- induced by the strong interactions of this
cationic impurity with helium~--
that 
suppresses the long exchange cycles and thus the superfluid response.
When accounting for the 
ro-vibrational
motion of
this molecule
however, the 
superfluid response of the first shell considerably increases.
These findings are consistent between the two different local estimators
of superfluidity as presented in the top and bottom panel of Fig.~\ref{fig:sfd}
as well as Table~\ref{tab:sf-shell}.
The main
difference between the two estimators is the 
extent 
of superfluidity,
where the exchange path estimator yields overall lower estimates throughout
compared to the area estimator.
By construction the exchange path length estimator will
always tend to underestimate the superfluid response while the area estimator
is expected to overestimate it, so that ultimately the ``real'' superfluid
density lies somewhere in between.
For further details and an 
in-depth
discussion we refer to SM Sec.~III.B.
\begin{table}[h]
   \centering
   \caption{Global superfluid fraction $f_\text{s}$ and local superfluid
           fractions associated with the first and second solvation
           shells of \meth{}$\cdot$He$_{60}$ and \pmeth{}$\cdot$He$_{60}$
           obtained by integrating the superfluid density computed
           using either the ``local area estimator''
           or the ``local exchange path estimator''
           with a cutoff length of $l=6$,
           see SM Sec.~III.B for details.}
   \renewcommand{\tabcolsep}{0.3cm}
   \renewcommand{\arraystretch}{0.62}
   \begin{tabular}{cccc}
      \toprule
      & \multicolumn{3}{c}{\meth{}$\cdot$He$_{60}$ -- Flexible molecule} \\
      \midrule
        & First shell & Second shell & Global \\
      Exchange path & 0.74 & 0.59 & 0.63\\
      Area & 0.96 & 0.83 & 0.86 \\
      \midrule
      &  \multicolumn{3}{c}{\meth{}$\cdot$He$_{60}$ -- Fixed molecule} \\
      \midrule
        & First shell & Second shell & Global \\
      Exchange path & 0.72 & 0.59 & 0.63 \\
      Area & 0.90 & 0.82 & 0.84 \\
      \midrule
      & \multicolumn{3}{c}{\pmeth{}$\cdot$He$_{60}$ -- Flexible molecule} \\
      \midrule
        & First shell & Second shell & Global \\
      Exchange path & 0.54 & 0.63 & 0.61 \\
      Area & 0.99 & 0.95 & 0.96 \\
      \midrule
      & \multicolumn{3}{c}{\pmeth{}$\cdot$He$_{60}$ -- Fixed molecule} \\
      \midrule
        & First shell & Second shell & Global \\
      Exchange path & 0.04 & 0.55 & 0.42\\
      Area & 0.28 & 0.83 & 0.68 \\
     \bottomrule
   \end{tabular}
   \label{tab:sf-shell}
\end{table}

Importantly, 
this 
enhancement of the superfluid response of the first shell is not 
due to a decrease of the local 
solid-like
order 
in the first shell as one could imagine, since the solvation 
structure 
is unaffected by the 
ro-vibrational
molecular motion of \pmeth{} as 
shown in
Fig.~\ref{fig:distrib}.
In a nutshell, in view of the strong localization of helium 
in the first shell around CH$_5^+$, one would expect almost no superfluid
response as obtained around
other strongly interacting 
impurities that lead to snowballs, and that is what is found when
fixing that molecule in its equilibrium structure.
However, including the full flexibility of CH$_5^+$, high
superfluid fractions are found, in particular in the first shell, 
thus showing that the superfluid response of helium is induced by 
the ro-vibrational motion of this flexible molecule
despite the pronounced localization of \he{}
close to this impurity.
We therefore conclude that 
helium 
in the first solvation shell around \pmeth{} 
features
\textit{manifestations of supersolid behavior} 
(in the spirit of \textit{``manifestations of superfluid behavior''}
predicted long ago for \textit{pure} \he{} clusters~\cite{sindzingre_path-integral_1989})
as 
indicated
by 
pronounced
bosonic exchange
in combination with
strong localization and
spatial order of \he. 
Evidently, the phenomenon 
uncovered here can only appear for clusters that are 
large enough to fully solvate the impurity and is thus 
absent in the microsolvation limit.
Are there any prospects to experimentally probe our prediction?
Manifestations of superfluid behavior in finite $^4$He~clusters 
were
first predicted based on seminal path integral 
simulations~\cite{sindzingre_path-integral_1989} and experimentally 
confirmed a decade later.
This experimental confirmation of \textit{``microscopic manifestation of superfluidity''} 
was made possible thanks to novel experimental ideas based on the IR spectra of a well-chosen 
molecular impurity used as probe of the local superfluidity in 
$^4$He~clusters~\cite{grebenev_superfluidity_1998}.
Concerning now manifestations of local supersolidity in \pmeth{}$\cdot$He$_n$ clusters, 
we refer to recent progress in measuring IR~spectra of various charged molecules in 
helium clusters and nanodroplets~\cite{ellis-cations-2019,vilesov-ch3+-2021,davies_infrared_2021,%
davies_onset_2023}, notably also including \pmeth{}. 
Similar to the original ideas behind 
what has been called the 
``microscopic Andronikashvili 
experiment''~\cite{grebenev_superfluidity_1998},
IR~spectra of \pmeth{} in $^3$He versus $^4$He nanodroplets as well as 
addition of a few $^4$He atoms to \pmeth{} in $^3$He could
reveal insightful differences.
Challenges certainly arise regarding IR spectroscopy of \pmeth{} since the molecule is known 
to stay fluxional in helium~\cite{uhl_helium_2018,uhl_quantum_2019,davies_infrared_2021} 
thus retaining its notorious spectroscopic complexity~\cite{asvany_understanding_2005}.
Exploring alternative observables different from IR~spectra might 
therefore 
provide complementary avenues for future experimental searches for this 
local supersolidity in doped He$^4$~clusters potentially based on experimental 
ideas that are yet to be developed.

\section{Conclusions and Outlook}
This study provides
strong evidence for manifestations of local supersolid behavior 
of \he{} around a charged molecular impurity, namely \pmeth. 
On the one hand, the strong interactions
between this molecular cation and helium 
lead to the 
long known
snowball effect, meaning a solid-like arrangement of helium in the first shell
with well-localized \he{} density. 
On the other hand, the 
strongly
localized atoms in this shell are 
involved in vivid bosonic exchange, induced by the 
ro-vibrational motion
of this fluxional molecular impurity,
in particular the
fast and 
complex rotational
motion emerging from the intimate coupling of overall rotations
and intramolecular pseudo-rotations. 
The combination of strong \he{} localization in the first shell 
with 
pronounced 
bosonic exchange 
therein, 
leading to an
intense superfluid response, thus clearly indicates 
manifestations of local supersolid behavior of \he{}
close to suitable molecular impurities~--
akin to the long-known manifestations of 
microscopic superfluidity in doped helium 
nanodroplets.
We expect this impurity-induced local supersolid
response
to appear in bosonic clusters doped with other impurities 
featuring a strong interaction with the solvent combined with
fast rotations resulting in a significant coupling between 
the impurity and the solvent.
In particular,
due to the 
stronger interactions of most 
impurities with para-H$_2$, 
we believe that the effect uncovered here 
could appear
as well in 
finite para-H$_2$ clusters 
doped with molecular impurities.
The phenomenon
is however markedly different
from the supersolidity that has long been predicted in pure
para-H$_2$ clusters of specific sizes, since the local
supersolid behavior is induced here by the molecular impurity
and is also not limited to magic numbers. 
Moreover, since the effect uncovered here can be expected to 
appear with various molecular ions in helium or 
other bosonic 
quantum fluids, it would be highly interesting
to explore in particular whether doping with 
an assembly of molecular ions 
serving as seeds
could lead to new supersolid phases 
especially in bulk helium or para-hydrogen.

\section{Acknowledgments}
We are thankful to Harald Forbert and Felix Uhl for many insightful
discussions.
This work was partially supported by DFG via MA~1547/19 and also 
funded
by the \textit{Deutsche Forschungsgemeinschaft}
(DFG, German Research Foundation) under Germany's
Excellence Strategy~-- EXC 2033~-- 390677874.
C.S. acknowledges partial financial support from the Alexander von
Humboldt-Stiftung and the 
DFG 
project number 500244608.
The computational resources were provided by HPC@ZEMOS,
HPC-RESOLV,
and BoViLab@RUB.
\end{document}

\title{Supplemental Material to \\Manifestations of Local Supersolidity of 
       $^\text{4}$He 
\\
around a Charged Molecular Impurity}
\author{Fabien Brieuc}
\email{fabien.brieuc@cea.fr}
\affiliation{Lehrstuhl f\"ur Theoretische Chemie,
 Ruhr-Universit\"at Bochum, 44780 Bochum, Germany}
\affiliation{Present address: Laboratoire Mati\`ere en Conditions Extr\^emes,
Universit\'e Paris-Saclay, CEA, 91680 Bruy\`eres-le-Ch\^atel; 
CEA, DAM, DIF, F-91297 Arpajon, France}
\author{Christoph Schran}
\email{christoph.schran@rub.de}
\affiliation{Lehrstuhl f\"ur Theoretische Chemie,
 Ruhr-Universit\"at Bochum, 44780 Bochum, Germany}
\affiliation{Present address: Yusuf Hamied Department of Chemistry,
University of Cambridge, Lensfield Road, Cambridge CB2 1EW, UK}
\author{Dominik Marx}
\email{dominik.marx@rub.de}
\affiliation{Lehrstuhl f\"ur Theoretische Chemie,
 Ruhr-Universit\"at Bochum, 44780 Bochum, Germany}
\date{\today}

\maketitle

\tableofcontents
\clearpage

\section{Neural Networks Potentials}
\label{sec:nnps} 
\subsection{Methodology}
High-dimensional NNPs~\cite{behler_generalized_2007,%
behler_first_2017} have been fitted against coupled cluster 
reference calculations including singles, doubles and perturbative 
triple excitations, CCSD(T). 
The augmented correlation-consistent basis set up to triple zeta
functions~\cite{%
kendall_electron_1992,woon_gaussian_1994} 
(aug-cc-pVTZ or AVTZ) 
has been used in combination
with the explicitly correlated 
F12a
method~\cite{adler_simple_2007,knizia_simplified_2009} 
using an adequate scaling of the triples~\cite{knizia_simplified_2009} 
(providing CCSD(T*)-F12a/AVTZ which is referred to simply as ``CC'' 
in the following), leading to a description of the electronic structure 
that is essentially converged to the complete basis set limit.
All reference calculations have been performed with the Molpro 
software package version 2012.1~\cite{werner_molpro_2012}.
The iterative procedure described in Refs.~\citenum{%
schran_high-dimensional_2017} and \citenum{schran_automated_2020} 
was used to optimally build the training sets of the models,
while ensuring that only a minimum number of reference calculations
were performed.
All atomic neural networks have a fully connected feed-forward 
architecture composed of two hidden layers of 25~nodes 
each
using a hyperbolic tangent as the activation function.
The output layer contains only one neuron activated by a linear
function as usual for regression.
Training of the models was performed with the RuNNer 
program~\cite{runner} using the element-decoupled Kalman filter 
optimizer~\cite{gastegger_high-dimensional_2015}.
The local atomic environments used as input of the models are 
described by atom-centered symmetry functions~\cite{behler_atom-centered_2011}.
We refer to the supporting file \texttt{nnp-parameters.pdf}
where all parameters of all neural network potentials
as used in this study are provided.

The interactions between the molecules and helium are described by 
separate NNPs that were fitted to reproduce the impurity{\ccdots}He
interaction in a pair-wise manner~\cite{schran_high-dimensional_2017}.
The reference interaction energies have been computed at the same level 
of theory as for the PESs of the molecules, but using the 
supermolecule approach and a counterpoise correction (cp) to correct for 
the basis set superposition error~\cite{boys_calculation_1970} 
(thus providing the CCSD(T*)-F12a-AVTZcp method, which we also refer to 
as ``CC'' for simplicity in what follows). 
The scaling correction of the triple excitations has been applied
in a system-size-consistent manner~\cite{marchetti_accurate_2009} 
by determining the correction only from the 
supermolecular cluster.
\subsection{Potential Energy Surface of CH$_5^+$} 
\begin{figure}
   \centering
    \includegraphics[width=0.8\textwidth]{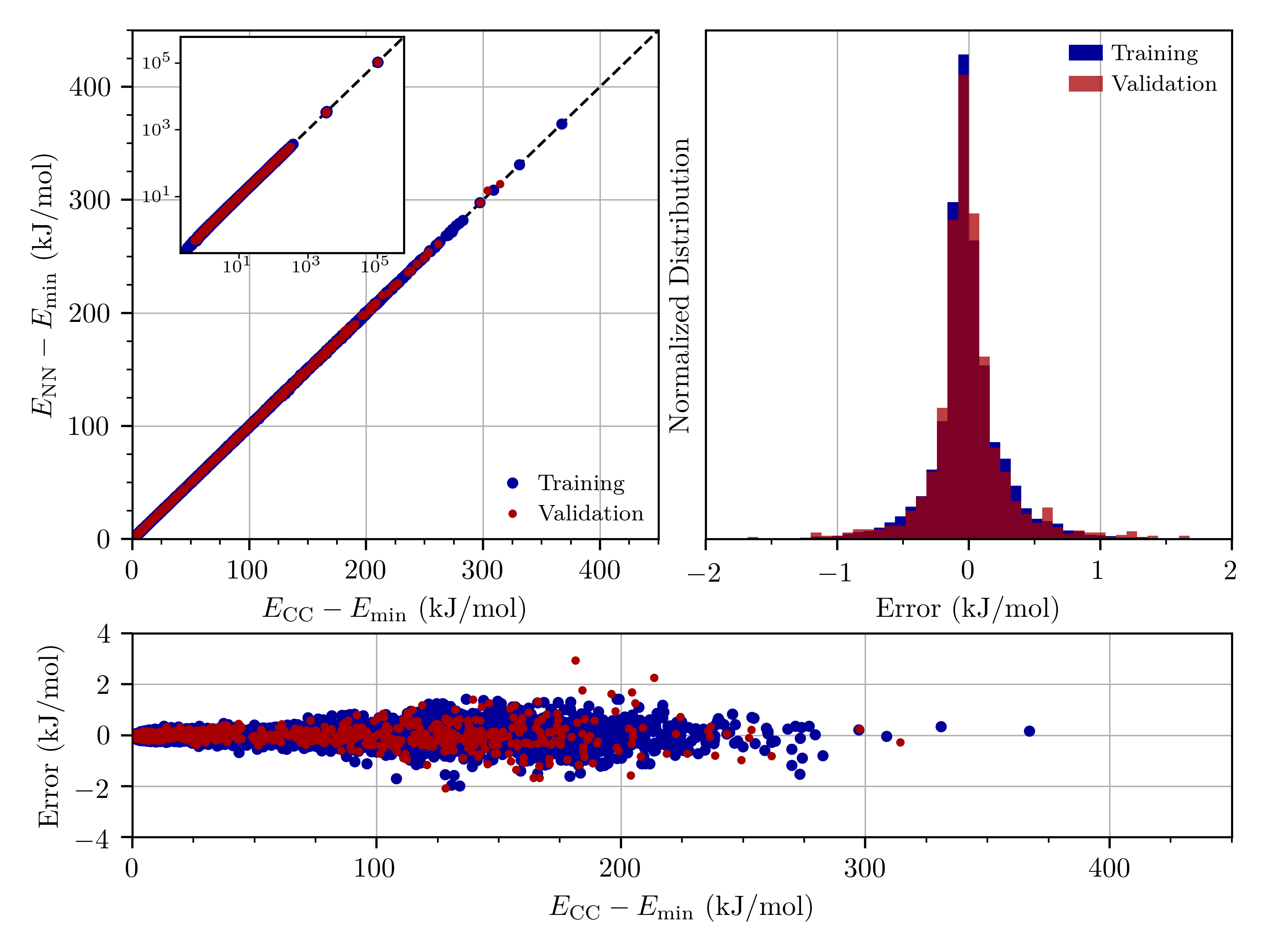}
   \caption{Analysis of the training of the \pmeth{} NN-PES.
      Correlations of the energies obtained by the NN-PES (NN) and the 
      reference coupled cluster (CC) method (top left). Histogram (top right) 
      and values (bottom) of the associated errors. The energies are reported 
      relative to the energy of the global minimum $E_\text{min}$.}
   \label{fig:ch5+-validation}
\end{figure}

\begin{figure}
   \centering
   \includegraphics[width=0.6\textwidth]{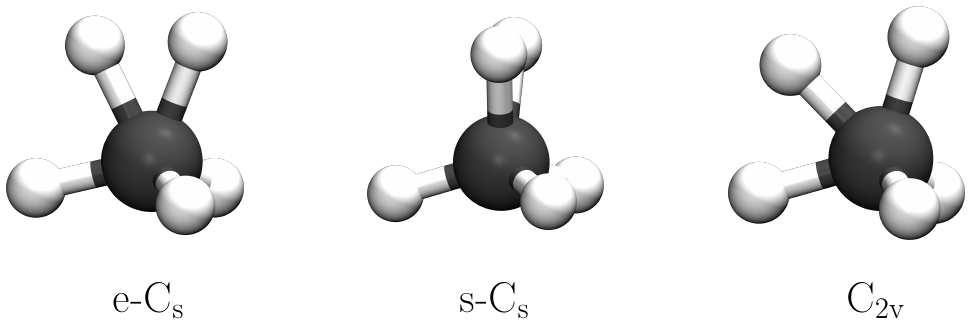}
   \caption{Important structures of \pmeth{}: minimum energy (left), 
            first stationary point (center) and second stationary
            point (right).}
   \label{fig:ch5+-struct}
\end{figure}

The training of the NNP to describe the PES
of \pmeth{}, 
thus providing what we call the NN-PES
(NN-PES-CH5P-2022-V0),
was performed on a training set composed of 18266~configurations
as sampled
by our automated fitting procedure from MD 
with classical nuclei and PIMD simulations considering nuclear quantum effects
at various temperatures ranging from~1 up to 1500\,K. 
A limited amount of \ce{CH3+} and \ce{H2} configurations were added in 
order to correctly describe configurations that are close to the dissociation 
channel (adding up to 800 \ce{CH3+} and 32 \ce{H2} representative structures).
Ten percent of the dataset is used as a validation set to assess the 
quality of the fit and to detect overfitting.
The root mean square error (RMSE) obtained for the training and 
validation set is 
0.3 and 0.4~kJ/mol, respectively, and the overall
high quality of the training is illustrated in Fig.~\ref{fig:ch5+-validation}.
The final NN-PES has been validated by investigating important stationary-point
structures of protonated methane represented by the minimum energy 
structure and the lowest two 
saddle points depicted in 
Fig.~\ref{fig:ch5+-struct}.
The global minimum energy structure of \pmeth{} has an eclipsed 
C$_{\rm s}$ (\eCs{}) point symmetry and is composed of a \ce{CH3} tripod 
and a \ce{H2} moiety connected through a three-center two-electron
bond~\cite{marx_structural_1995}.
This \ce{H2} moiety can rotate around the C$_3$ axis of the tripod, 
a motion associated to a stationary point with staggered C$_{\rm s}$ 
(\sCs{}) symmetry and a small energy barrier that makes 
this pseudo-rotation essentially barrier-free.
The second low-lying stationary point has the \Ctv{} symmetry and 
is associated with another pseudo-rotation that connects two 
degenerate \eCs{} structures involving different atoms in the moiety.
The combination of these two pseudo-rotations leads to a large-amplitude 
motion, sometimes called scrambling motion, that makes all the hydrogen 
atoms dynamically equivalent.
Since this scrambling motion is associated with tiny energy barriers 
that are overcome by zero-point energy, \pmeth{} exhibits large-amplitude 
motion even in the $T \to 0$ limit: Protonated methane is the typical 
example of a quantum fluxional molecule.

Geometry optimizations of these different structures of \pmeth{}
have been performed using both, the NN-PES and explicit CC reference
calculations.
Their
relative energies are compared in Table~\ref{tab:ch5+-dE} in order to 
confirm that the energy barriers associated with the large amplitude motion 
of protonated methane are correctly reproduced by the NN-PES. 
In particular, these
tests show that the NN-PES represents the stationary points of \pmeth{}
on its complex PES to very high precision
which greatly exceeds ``chemical accuracy''. 
Moreover, the normal mode frequencies of \pmeth{} in the minimum 
energy structure have been computed with the NN-PES and 
are compared in Table~\ref{tab:ch5+-NM} to the values obtained with 
the CC reference method.
Also here, the NN-PES is able to reproduce the CC reference within a few~cm$^{-1}$, 
highlighting the quality of this approach.
\begin{table}
   \centering
   \renewcommand{\tabcolsep}{0.3cm}
  \caption{Energy difference $\Delta E (X) = E(X) - E(\text{e-C}_\text{s})$ 
           in kJ/mol as obtained using the NN-PES and the reference CC
           method with associated errors.}
   \label{tab:ch5+-dE}
   \begin{tabular}{ccccc}
      \toprule
      $X$ & NN & CC & Error \\
      \midrule
      \sCs{} & 0.3855 & 0.4296 & 0.0441 \\
      \Ctv{} & 3.817  & 3.6559 & $-0.1611$  \\
     \bottomrule
  \end{tabular}
\end{table}
\begin{table}
   \centering
   \renewcommand{\tabcolsep}{0.15cm}
   \caption{Normal mode frequencies in cm$^{-1}$ in the minimum energy 
            \eCs{} structure of \pmeth{} as obtained using the NN-PES 
            and the reference CC method with associated errors. The mean
            absolute deviation (MAD) is also reported on the last line.}
   \label{tab:ch5+-NM}
   \resizebox{1.0\textwidth}{!}{
      \begin{tabular}{ccccccccccccc}
         \toprule
       NN    & 226.48 & 831.19 & 1284.52 & 1301.37 & 1460.18 & 1482.71 & 1590.29 & 2436.55 & 2720.25 & 3004.74 & 3135.13 & 3238.15 \\
       CC    & 211.92 & 828.23 & 1282.81 & 1289.30 & 1461.79 & 1490.27 & 1595.63 & 2449.88 & 2713.51 & 3014.75 & 3133.99 & 3228.55 \\
       Error & $-14.56$ &  $-2.96$ &  $-1.71$ &  $-12.1$ &  $1.61$ &  $7.56$ &  $5.34$ &  $13.3$ & $-6.74$ &  $10.0$ & $-1.14$ & $-9.60$ \\
       MAD   & 7.2      &          &          &          &         &         &         &         &         &         &         &      \\
         \bottomrule
      \end{tabular}
   }
\end{table}

\begin{figure}
   \centering
   \includegraphics[width=0.8\textwidth]{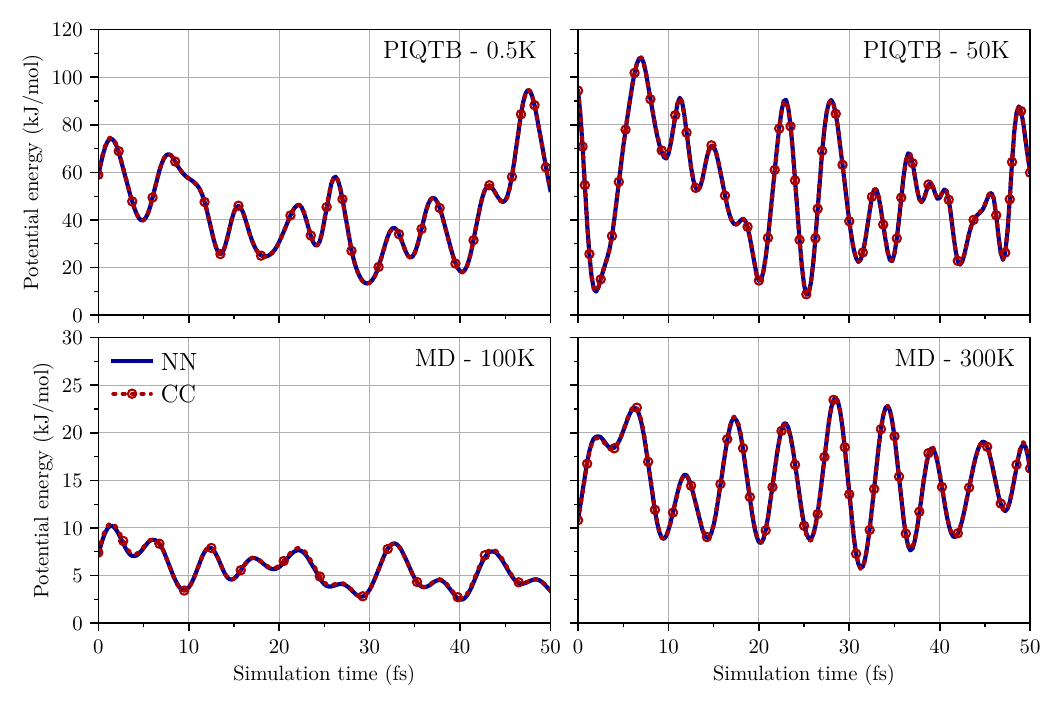}
   \caption{Potential energy of an isolated \pmeth{} molecule along one 
            replica of PIQTB trajectories (first row) at $T=0.5$\,K (left) 
            and 50\,K (right) and along a standard MD trajectory (second row) 
            at $T=100$\,K (left) and 300\,K (right). The coupled cluster 
            reference data (CC) were obtained by recomputing the energies 
            at each step of the NN-PES trajectories and are shown as red 
            dashed lines (with only a few circles added since the CC 
            energies practically superimpose the NN-PES data). All energies 
            are reported relative to the energy of the global minimum (\eCs{}).}
   \label{fig:ch5+-traj}
\end{figure}

Finally, in order to test further the NN-PES on structures actually sampled 
in realistic simulations, we randomly extracted short pieces of MD and PIMD 
trajectories of a single \pmeth{} in the gas phase in which the potential 
energy was evaluated using both the NN-PES and the CC reference method as 
depicted in Fig.~\ref{fig:ch5+-traj}.
Independent of whether the nuclei are described as classical point particles 
or including nuclear quantum effects, the NN-PES provides 
perfect agreement
with the CC reference energies 
covering 
very different temperature regimes. 
These various tests clearly highlight the 
outstanding
quality of this NN-PES 
of \pmeth{} that essentially allows for a description of the molecule at 
converged coupled cluster accuracy.
\subsection{Potential Energy Surface of CH$_4$} 
\begin{figure}
   \centering
   \includegraphics[width=0.8\textwidth]{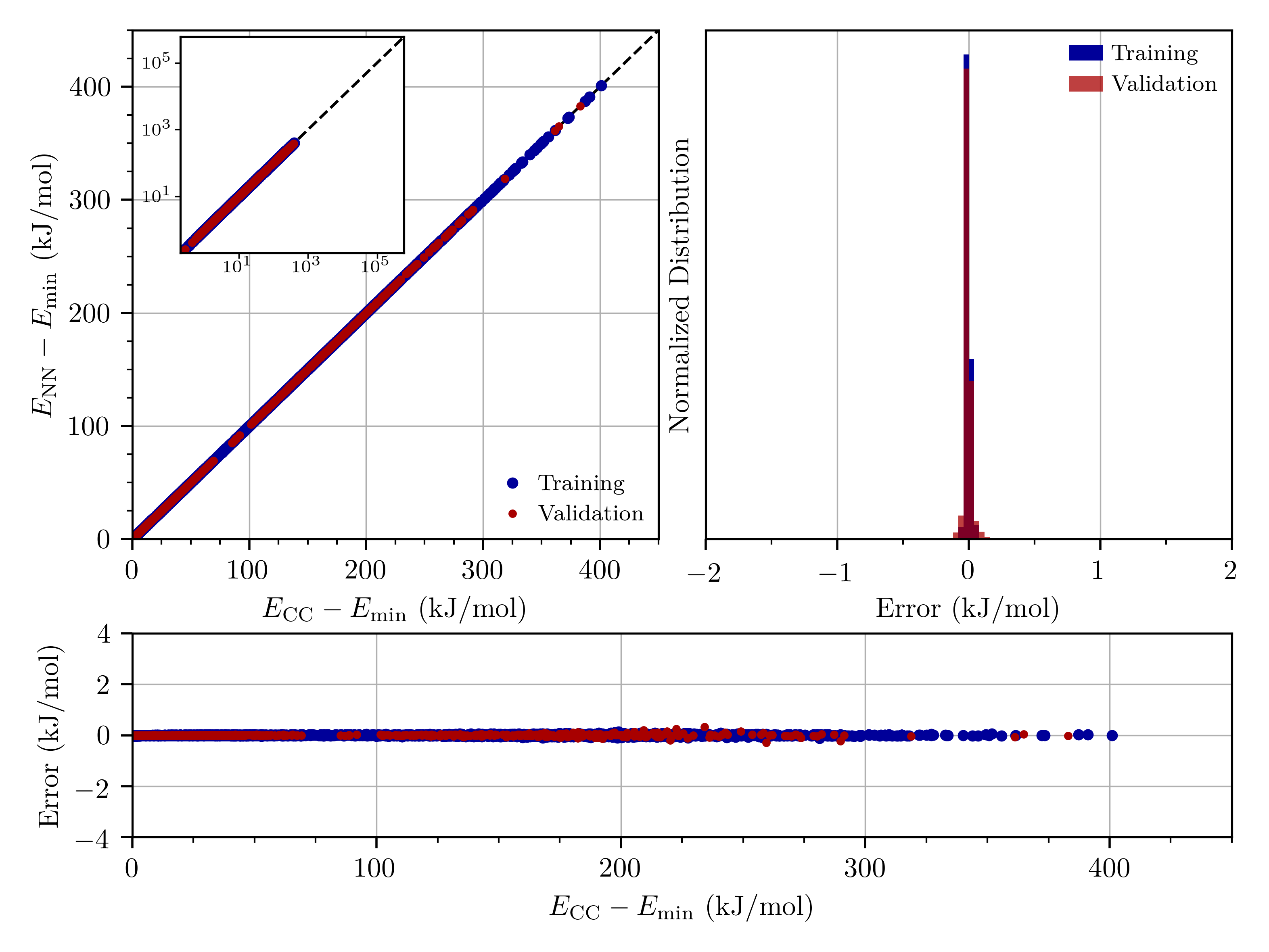}
   \caption{Analysis of the training of the \meth{} NN-PES.
      Correlations of the energies obtained by the NN-PES (NN) and the 
      reference coupled cluster (CC) method (top left). 
      Histogram (top right) and values (bottom) of the associated errors. 
      The energies are reported relative to the energy of the global 
      minimum $E_\text{min}$. We employ here on purpose the same scale 
      as used for the validation of the \pmeth{} NN-PES 
analyzed in Fig.~\ref{fig:ch5+-validation} 
      in order to allow for one-to-one comparison of this two cases
which represent quasi-rigid (\meth{}) and highly fluxional (\pmeth{}) molecules
of very similar kind and size.}
   \label{fig:ch4-validation}
\end{figure}

\begin{figure}[H]
   \centering
   \includegraphics[width=0.8\textwidth]{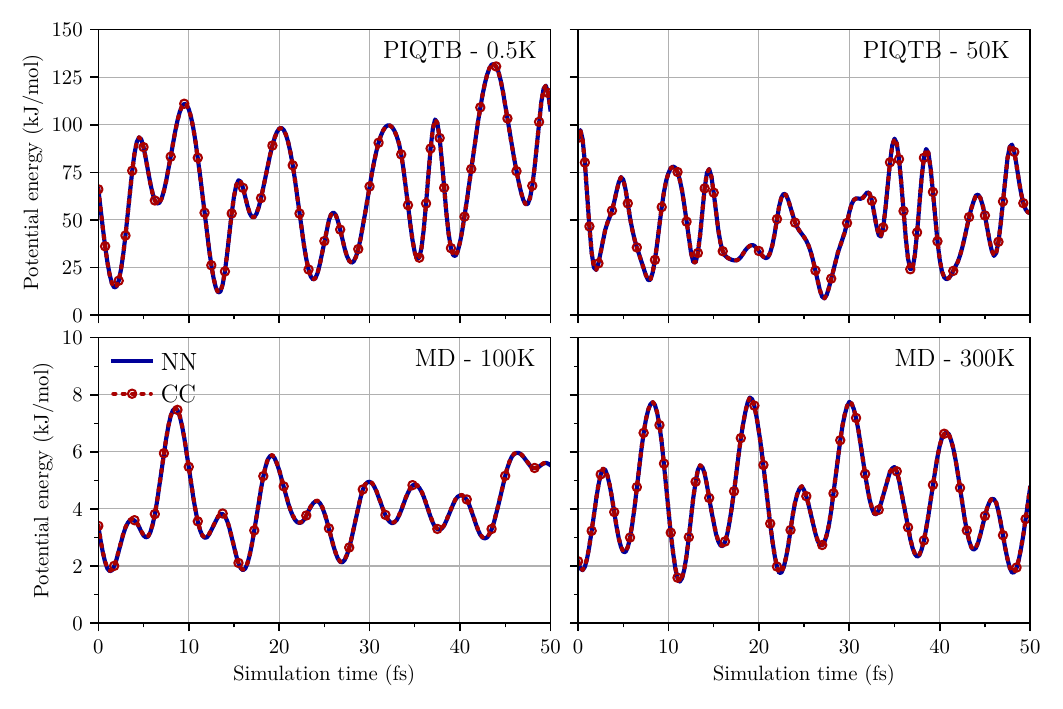}
   \caption{Potential energy of an isolated \meth{} molecule along one 
            replica of PIQTB trajectories (first row) at $T=0.5$\,K 
            (left) and 50\,K (right) and along a standard MD trajectory 
            (second row) at $T=100$\,K (left) and 300\,K. The coupled 
            cluster reference data (CC) were obtained by recomputing the 
            energies at each step of the NN-PES trajectories and are shown 
            as red dashed lines (with only a few circles added since the 
            CC energies practically superimpose the NN-PES data). All 
            energies are reported relative to the energy of the global 
            minimum.}
   \label{fig:ch4-traj}
\end{figure}

The NN-PES of \meth{} 
(NN-PES-CH4-2022-V0)
has been developed following the same automated 
fitting procedure as used for \pmeth{}.
The dataset generated here is composed of 9245 configurations which have 
been selected from both PIMD and MD trajectories of isolated \meth{} at 
temperatures ranging from 1.67 up to 1000\,K.
The RMSE associated with the training of the NN-PES are 
0.02~kJ/mol
the training set and 
0.03~kJ/mol for the validation set.
The quality of the fit is further illustrated in
Fig.~\ref{fig:ch4-validation}
at the level of errors, which 
clearly highlights 
that fitting the NN-PES of such a quasi-rigid and symmetric 
molecule as \meth{} is a much easier task than fitting the 
complex PES of highly fluxional molecules such as \pmeth{}.
\begin{table}[b]
   \centering
   \renewcommand{\tabcolsep}{0.25cm}
   \caption{Normal mode frequencies in cm$^{-1}$ in the minimum energy 
            structure of \meth{} as obtained using the NN-PES 
            and the reference CC method with associated errors. The mean
            absolute deviation (MAD) is also reported on the last line.}
   \label{tab:ch4-NM}
      \begin{tabular}{ccccccccccccc}
         \toprule
         NN    & 1344.04 &  1344.04 &  1344.04 &  1568.65 &  1568.65 &  3036.86 &  3156.00 &  3156.00 & 3156.00 \\
         CC    & 1343.20 &  1343.32 &  1343.37 &  1568.09 &  1568.20 &  3032.68 &  3155.17 &  3155.26 & 3155.36 \\
         Error & $-0.84$ &  $-0.72$ &  $-0.67$ &  $-0.56$ &  $-0.45$ &  $-4.18$ &  $-0.83$ &  $-0.78$ & $-0.64$ \\
         MAD   & 1.1   &          &          &          &          &          &          &          &   \\      
         \bottomrule
      \end{tabular}
\end{table}
The minimum energy structure is very well reproduced with an error on the 
energy that is lower than 0.05~kJ/mol and errors on the associated 
normal modes of around 1~cm$^{-1}$ only, as compiled in Table~\ref{tab:ch4-NM}.
As for \pmeth{}, the quality of the NN-PES is further tested on short 
parts of representative MD and PIMD trajectories obtained using the NN-PES
for which the potential energy gets re-evaluated using the reference 
CC  method, see Fig.~\ref{fig:ch4-traj}.
For all these validation simulations~-- including various temperatures and both
classical and quantum nuclei~--  essentially perfect agreement
between the NN-PES and the CC reference is observed.
\subsection{CH$_5^+$${\ccdots}$Helium Interaction}
\begin{figure}[h]
   \centering
   \includegraphics[width=0.7\textwidth]{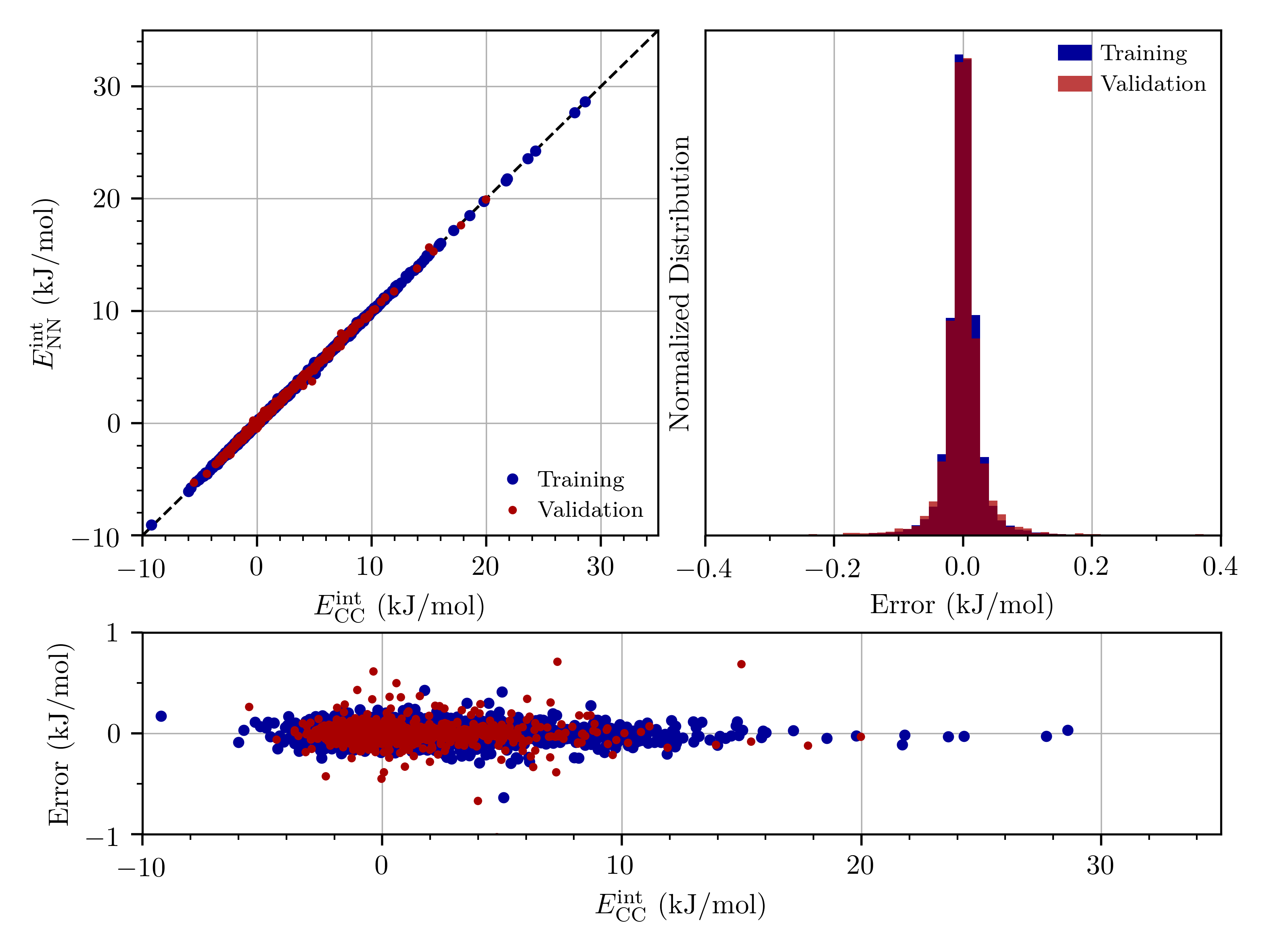}
   \caption{Analysis of the training of the \pmeth{}$\ccdots$\he{}
      NN-PES. Correlations of the interaction 
      energies obtained by the NN-PES (NN) and the reference coupled cluster 
      (CC) method (top left). Histogram (top right) and values (bottom) of 
      the associated errors.}
   \label{fig:ch5+-he-validation}
\end{figure}

\begin{figure}[h]
   \centering
   \includegraphics[width=0.75\textwidth]{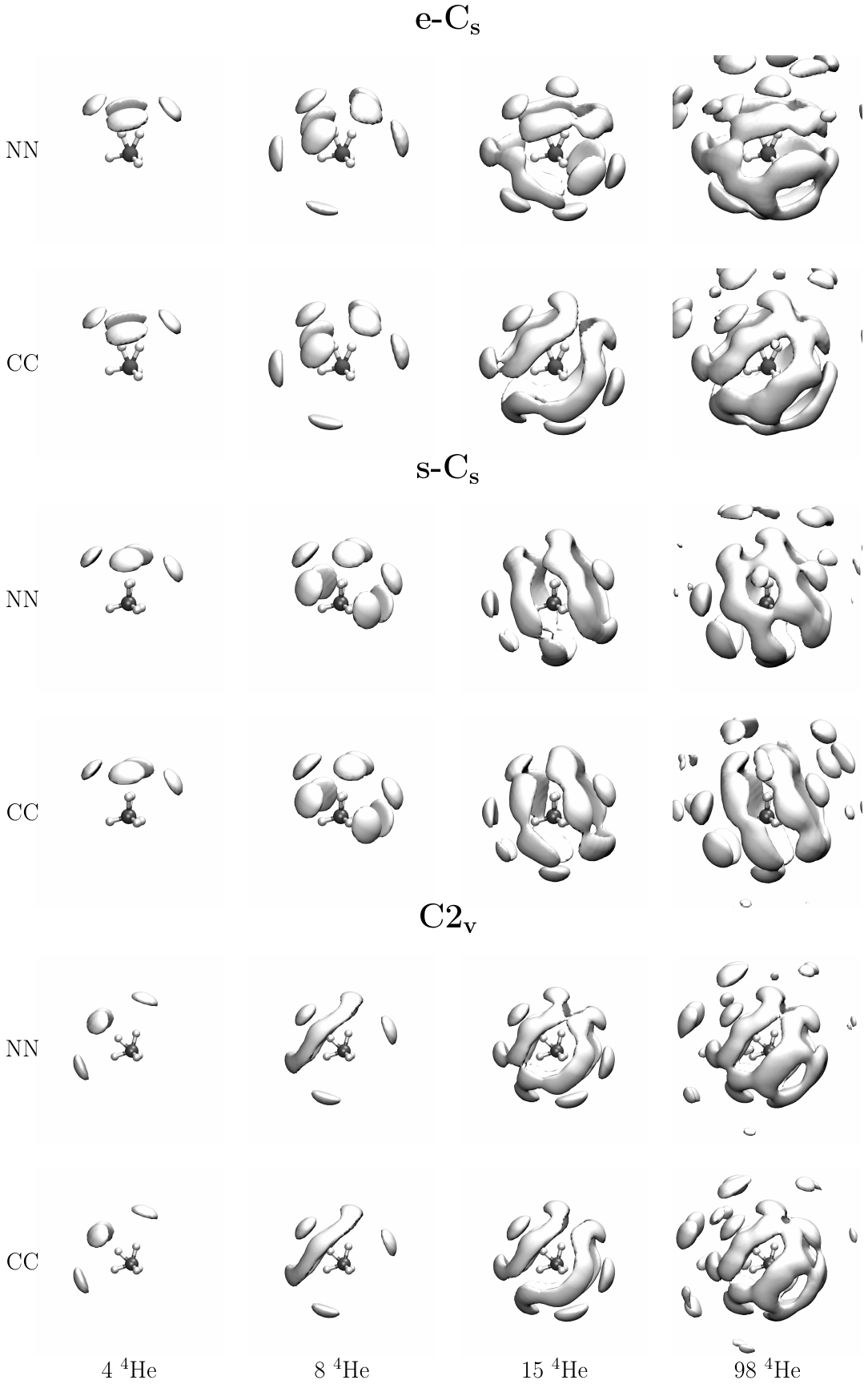}
   \caption{Helium densities (spatial distribution functions) 
      around \pmeth{} completely fixed in space 
      in various representative configurations, \eCs{} (two first 
      rows), \sCs{} (third and fourth rows) and \Ctv{} (two last rows).
      The densities are computed by PIMC at $T=1.67$\,K 
      without bosonic exchange and using potential 
      energy values that have been precomputed with the NN-PES
      (NN) and the reference coupled cluster method (CC) 
      on a fine grid around the fixed molecule.
      Isosurface values: 0.035 $N_\text{He}$/\AA$^3$ for 4 He, 
      0.025 $N_\text{He}$/\AA$^3$ for 8 He, 0.020 $N_\text{He}$/\AA$^3$ 
      for 15 He and 0.015 $N_\text{He}$/\AA$^3$ for 98 He.}
   \label{fig:ch5+-he-densities}
\end{figure}

The \pmeth{}$\ccdots$\he{} interaction has been 
parameterized in terms of a NN-PES
(NN-IP-CH5P-HE-2022-V0)
using another automated procedure as
described for the first time in Ref.~\citenum{schran_high-dimensional_2017},
which is similar to the one used for fitting the molecular PES
but deviating in crucial aspects as explained previously. 
The dataset used for this purpose is composed of 48000 \pmeth{}$\cdot$\he{} 
configurations.
The RMSE associated with the training and validation set is 
0.03 and 
0.05~kJ/mol, respectively, and the quality of the 
fit is illustrated in Fig.~\ref{fig:ch5+-he-validation}.
To further test this neural network description of the 
\pmeth{}$\ccdots$\he{} interactions, we put it to a most stringent 
test by directly comparing the helium density 
(in terms of its spatial distribution function, SDF)
obtained using the
NN-PES and the CC reference method around a space-fixed molecular impurity.
In practice, the interaction potential is first tabulated on 
a fine grid around the 
fixed 
molecule: For each point, the 
interaction energy is computed using both the NN-PES and the 
CC method.
PIMC simulations of helium around 
the molecule are then performed 
using the precomputed energy value on the closest grid point.
This allows us to compute approximate helium densities in Cartesian 
space around a fixed molecular impurity~-- even 
with the computationally expensive CC method.
The results of this comparison are presented in 
Fig.~\ref{fig:ch5+-he-densities} and clearly show that the NN-PES
correctly reproduces the helium densities around \pmeth{} in 
all important 
configurations,
namely \eCs{}, \sCs{} and \Ctv{}, 
even for a large number of helium atoms (98 atoms here).
Upon closer inspection, 
one can detect
some small remaining differences 
for large numbers of 
\he{} atoms 
(where smaller isovalues are required to visualize the larger clusters)
around \pmeth{} in the \eCs{} configuration;
a similar observation has been made and analyzed 
earlier~\cite{schran_high-dimensional_2017} in Fig.s~3 and~4 therein. 
Nevertheless, the most prominent features of the density remain 
correctly described even for 98~helium atoms and the NN-PES, thus,
still provides  
an accurate 
description of the helium density in these cases.
\clearpage
\subsection{CH$_4${\ccdots}Helium Interaction}
\begin{figure}[b]
   \centering
   \includegraphics[width=0.8\textwidth]{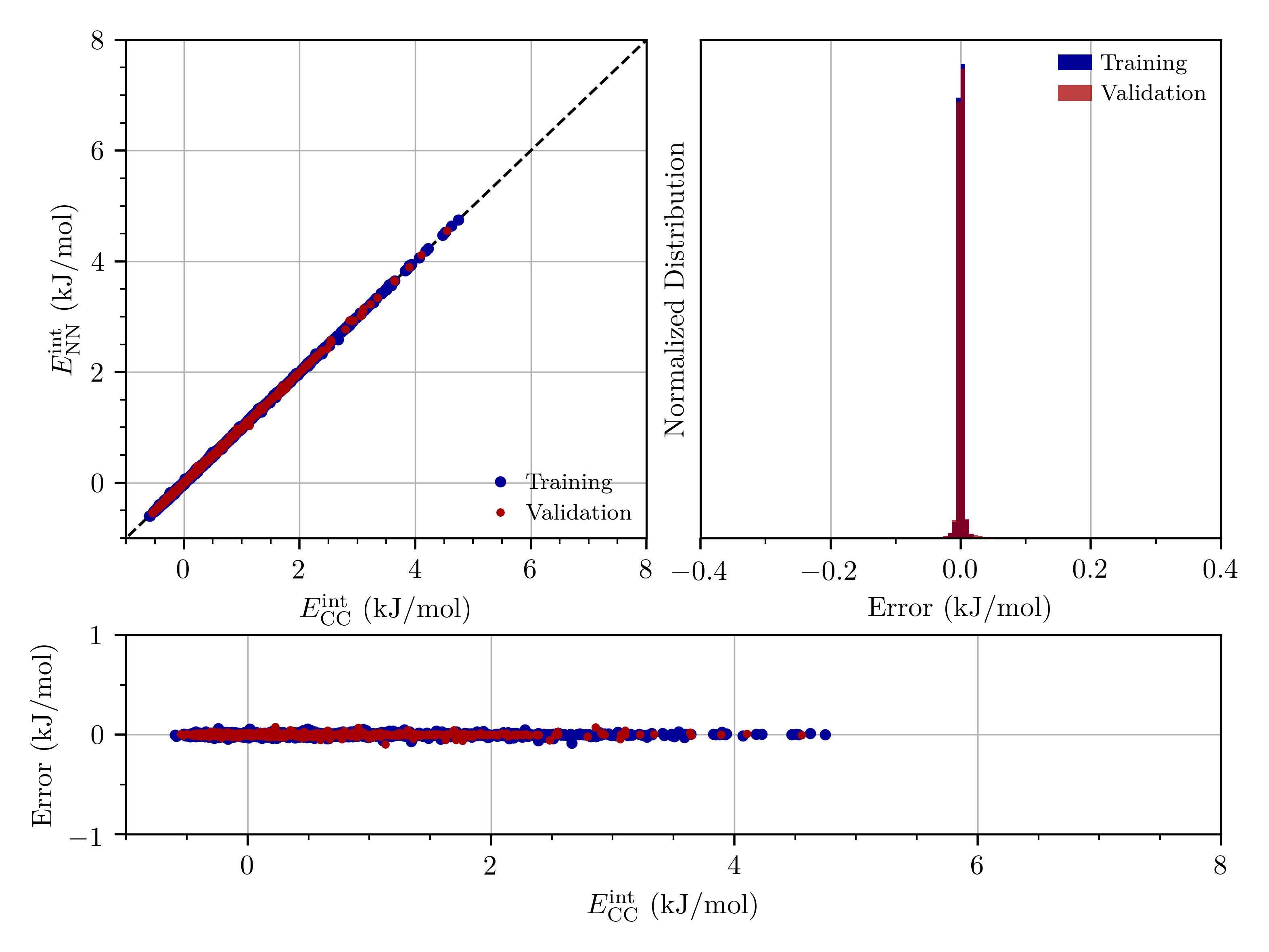}
   \caption{Analysis of the training of the \meth{}$\ccdots$\he{}
      NN-PES. Correlations of the interaction
      energies obtained by the NN-PES (NN) and the reference coupled cluster
      (CC) method (top left). Histogram (top right) and values (bottom) 
      of the associated errors.}
   \label{fig:ch4-he-validation}
\end{figure}

\begin{figure}
   \centering
   \includegraphics[width=0.8\textwidth]{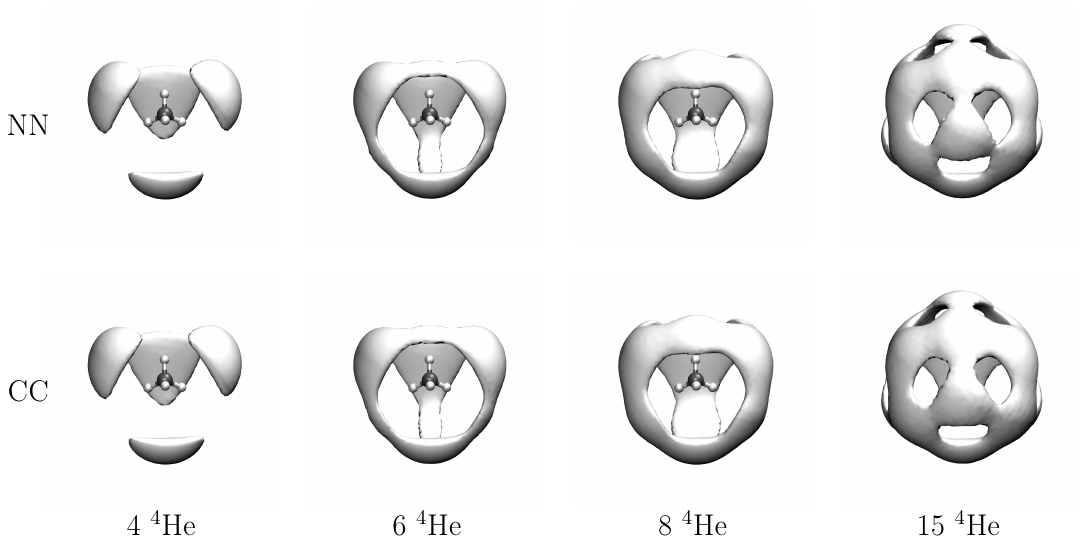}
   \caption{Helium densities (spatial distribution functions)
      around \meth{} completely fixed in space in 
      its minimum energy configuration. The densities are computed 
      by PIMC at $T=1.67$\,K without bosonic exchange and using 
      potential energy values that have been precomputed with the NN-PES
      (NN) and the reference coupled 
      luster method (CC) on a fine grid around the fixed 
      molecule.
      Isosurface values: 0.0040 $N_\text{He}$/\AA$^3$ for 4 and 6 He, 
      0.0050 $N_\text{He}$/\AA$^3$ for 8 He and 0.0055 
      $N_\text{He}$/\AA$^3$ for 15 He.}
      \label{fig:ch4-he-densities}
\end{figure}

As for \pmeth{}$\ccdots$\he{}, we use a dedicated high-dimensional NNP to describe the 
\meth{}$\ccdots$\he{} interaction potential 
(NN-IP-CH4-HE-2022-V0).
The generation of the dataset as well as the training of the interaction potential
are performed using the same automated approach as before.
The dataset is composed of 47833 \meth{}$\cdot$\he{} configurations
and the RMSE associated with the training and validation set is 
0.004 and 
0.005~kJ/mol, respectively.
The quality of the fit is illustrated in Fig.~\ref{fig:ch4-he-validation},
which clearly shows that the NN-PES is able to reproduce the CC reference
to very high precision.
To further test this description of the \meth{}$\ccdots$\he{} interaction,
we put it to the same test as before and directly compare the helium densities
obtained using the NN-PES and the CC reference method around a fixed 
molecular impurity in the minimum energy configuration.
The results presented in Fig.~\ref{fig:ch4-he-densities} clearly show that
this approach is again able to describe the helium solvation of the molecule with 
great accuracy
since essentially no difference between the densities can be observed.

\newpage
\section{Solvation Shell Structure}
\begin{figure}[bh!]
   \centering
   \includegraphics[width=\textwidth]{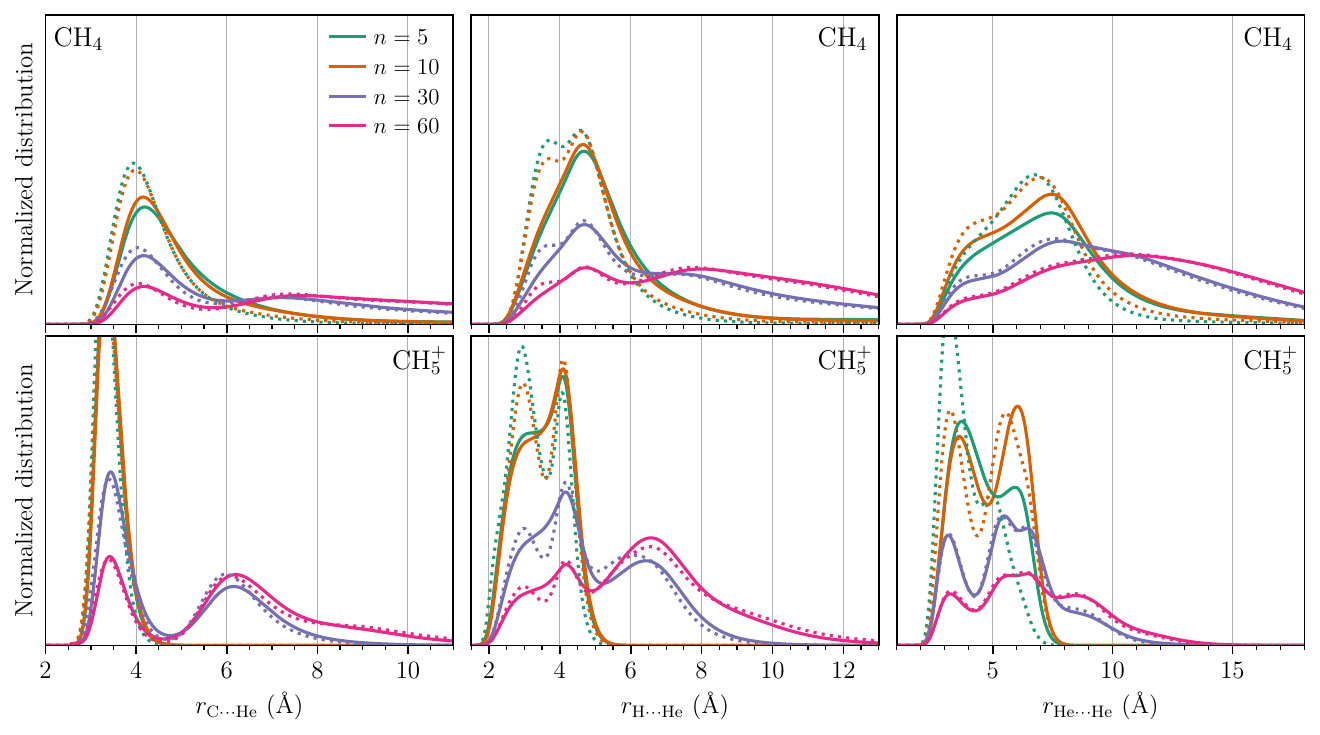}
   \caption{Structure of the helium solvation shell around \meth{} 
            (first row) and \pmeth{} (second row) for different numbers $n$ of 
            \he{} atoms at $T=0.5$\,K. 
The dotted lines correspond to molecules completely fixed in space in 
their minimum energy structures. 
The first column presents distributions of 
            C$\ccdots$He~distances, the second column presents distributions 
            of H$\ccdots$He~distances and the last column presents 
            distributions of He$\ccdots$He~distances. 
            All the distributions are normalized to unity.}
   \label{fig:he-distrib}
\end{figure}

\begin{figure}[th!]
   \centering
   \includegraphics[width=0.65\textwidth]{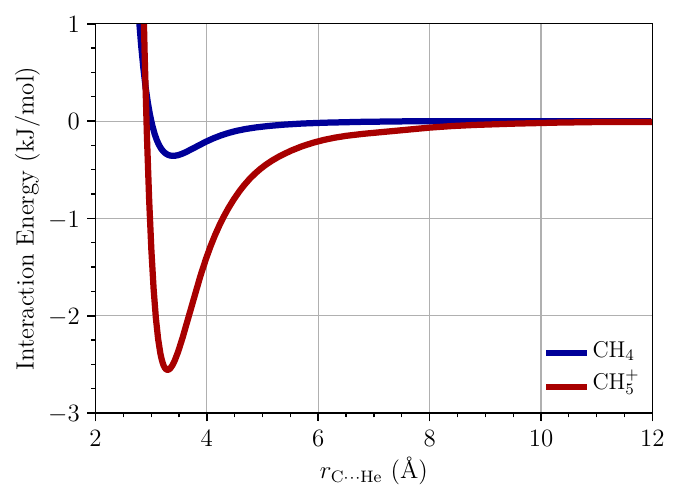}
   \caption{Interaction energy as a function of the C$\ccdots$He distance 
            along the direction of maximum interaction for both 
            \meth{} and \pmeth{} fixed in their minimum energy structure.}
   \label{fig:scan}
\end{figure}

\begin{figure}[bh!]
   \centering
   \includegraphics[width=0.8\textwidth]{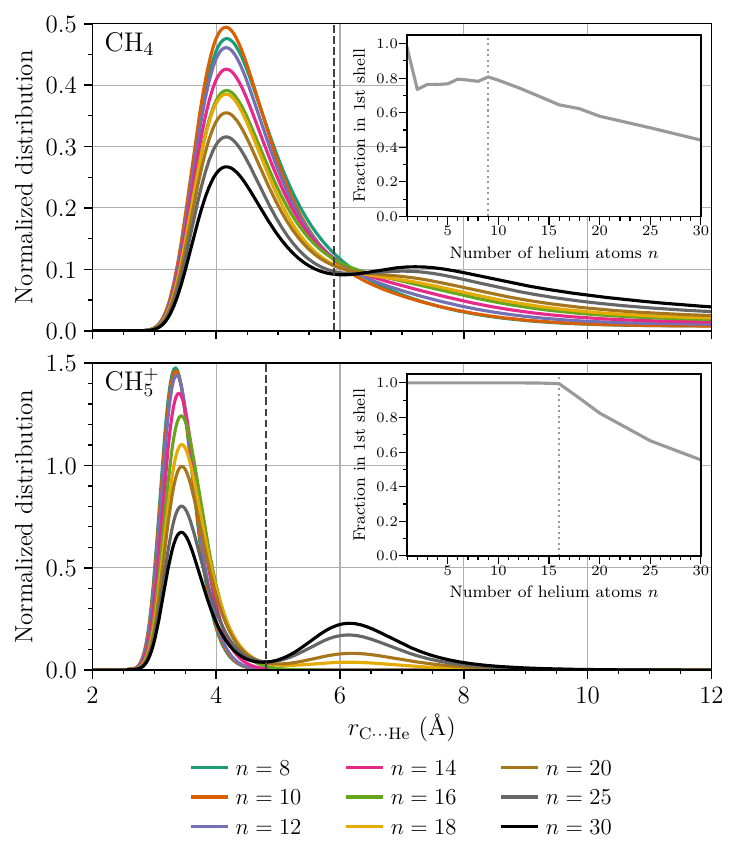}
   \caption{Distribution of  C$\ccdots$He distances around \meth{} (top)
            and \pmeth{} (bottom) at $T=0.5$\,K for different numbers of 
            helium atoms $n$
as provided below the plots.
            The insets show the fraction of helium
            atoms that are in the first shell as a function of the 
            number of \he{} atoms $n$ computed by integrating
            the density in the first shell;
the vertical dotted lines
            indicate 
the $n$ values 
when the second shell starts to be gradually filled.
            The distance used to separate between first and second shell 
            is 5.9~\AA{} for \meth{} and 4.8~\AA{} for \pmeth{} as marked 
            by the dashed vertical lines in the main plots.}
   \label{fig:shell-filling}
\end{figure}

In addition to the structural properties presented in the main text,
we present here further details of
the structure of the helium solvation shell.
Various distributions are presented in Fig.~\ref{fig:he-distrib},
which show that helium around \pmeth{} is 
much more structured than around \meth{}, an effect that is the  
direct result of the much stronger interaction between helium and 
protonated methane as quantified in Fig.~\ref{fig:scan}.
One can also see from the distributions in Fig.~\ref{fig:he-distrib} 
that completely neglecting the molecular degrees of freedom 
by fixing the two molecules at their minimum energy configurations
does not qualitatively change the helium solvation shell structure
for large clusters. 
In particular, in the large cluster limit, $n=60$, the impact of molecular motion 
on the overall structure of the solvation shell is essentially negligible.
Focusing more closely on the C$\ccdots$He distance distributions, 
a clear shell structure appears around \pmeth{} 
featuring a sharp first solvation shell localized between~2.5 and 
4.8~\AA{}, which is filled with~16 \he{} atoms as analyzed in 
detail in Fig.~\ref{fig:shell-filling}.
In the case of \meth{}, however, only a faint shell structure is observed 
with a first shell that is much less clearly defined and located between 
around~2.75 and 5.9~\AA{}.
Studying the evolution of the C$\ccdots$He distance as a function
of the number of helium atoms $n$ 
in Fig.~\ref{fig:shell-filling}
shows that a second shell is formed around \pmeth{} for 
$n>16$ \he{} atoms, 
see in particular the corresponding inset. 
The situation is, however, less clear around \meth{} where a second 
shell can only be hinted for $n>20$ \he{} atoms.
When looking at the fraction of atoms that form
the first shell it
is evident
that the second shell already starts
to be gradually filled for $n>9$ \he{} atoms
around \meth{},
as seen in the inset of Fig.~\ref{fig:shell-filling},
whereas shell filling is sharp in the case of the strongly 
interacting  \pmeth{} molecule. 

\FloatBarrier
\section{Superfluid Estimators}
\subsection{Superfluid Fraction}
\label{sec:fs} 
Within the two fluid model of superfluidity, the density of
helium $\rho$ is divided into a superfluid component $\rho_\text{s}$ 
and a normal component $\rho_\text{n}$ with
\begin{equation}
   \rho = \rho_\text{n} + \rho_\text{s}.
\end{equation}
The superfluid fraction 
can then be 
defined as the fraction of helium 
in the superfluid state $f_\text{s}=\rho_\text{s}/\rho$ and is 
connected to the decrease of the effective moment of inertia 
$I^\text{eff}$ that appears at temperatures lower than the critical 
temperature $T_\text{c}$ of the superfluid transition
\begin{eqnarray}
   f_\text{s} & = 1 - f_\text{n} = 1 - \frac{I^\text{eff}(T<T_\text{c})}{I^\text{eff}(T=T_\text{c})},
   \label{eq:fs-1}
\end{eqnarray}
with $f_\text{n} = \rho_\text{n}/\rho$ being the normal fraction.
This formula can of course only be used in cases where a 
proper superfluid phase transition appears for which a critical temperature 
can unambiguously be defined, as done experimentally
for bulk \he{} at ambient pressure (while assuming that 
the total density does not change as a function of temperature in that range). 
In this case, $T_\text{c}=2.17$\,K is the standard critical 
temperature of the lambda transition.

In the famous Andronikashvili experiment~\cite{andro_1946} the effective moment of 
inertia is measured through the angular frequency of a torsional 
oscillator immersed in liquid helium.
Upon decreasing the temperature, the contribution of the helium to 
the total moment of inertia, i.e. the effective moment of 
inertia, decreases and reaches zero when the helium is entirely 
superfluid ($f_{\rm s}$=1).
In simulations, the effective moment of inertia can be obtained 
using the following relation from linear response theory
\begin{equation}
I^\text{eff}_\alpha = \frac{\partial \langle\hat{L}_\alpha\rangle}{\partial \omega}\bigg|_{\omega=0}
\label{eq:Ieff-1}
\end{equation}
that connects the $\alpha$ principal component of the effective moment of inertia  
to the response of helium to an infinitesimally slow rotation around
the principal $\vec{\text{e}}_\alpha$ axis;
note that any axis can be taken in the homogeneous isotropic bulk limit. 
Within the path integral framework, Eq.~(\ref{eq:Ieff-1}) can be used 
to obtain the following expression for the effective moment of inertia
\begin{equation}
   I^\text{eff}_\alpha = I^\text{cl}_\alpha - 
                         \frac{4m_\text{He}^2}{\beta\hbar^2} \langle A_\alpha^2 \rangle,
   \label{eq:Ieff-2}
\end{equation}
with $I^\text{cl}_{\alpha}$ the $\alpha$ principal component of the classical 
moment of inertia which can be expressed as
\begin{equation}
   I^\text{cl}_\alpha = \left\langle \frac{m_\text{He}}{P} \sum_{i=1}^{N_\text{He}} \sum_{s=1}^P 
   (\vec{\text{e}}_\alpha \times \vec{r}_{i,s}) \cdot (\vec{\text{e}}_\alpha \times \vec{r}_{i,s+1}) \right\rangle
\end{equation}
and $A_\alpha^2$ the square of the vectorial area of the Feynman paths
\begin{equation}
   \vec{A} = \frac{1}{2}\sum_{i=1}^{N_\text{He}}\sum_{s=1}^P \vec{r}_{i,s} \times \vec{r}_{i,s+1},
\end{equation}
i.e. the area spanned by the ring polymer, projected along 
the $\vec{\text{e}}_\alpha$ direction.
The second term on the right hand side of equation~(\ref{eq:Ieff-2}) 
quantifies the quantum reduction of the effective moment of inertia due to 
superfluidity.
The superfluid fraction is thus obtained in a similar 
way as in equation~(\ref{eq:fs-1}) 
\begin{equation}
   f_{\rm s}^\alpha = 1 - \frac{I^\text{eff}_\alpha}{I^\text{cl}_\alpha} 
              = \frac{4m_\text{He}^2\langle A_\alpha^2 \rangle}{\beta\hbar^2I^\text{cl}_\alpha}.
   \label{eq:fs-area}
\end{equation}
This estimator is referred to as the ``area estimator'' and has been 
first introduced
for the study of pure
helium clusters~\cite{sindzingre_path-integral_1989}; 
a detailed discussion and derivation can be found in     
Ref.~\citenum{zeng_microscopic_2014}.

Obviously, superfluidity is a macroscopic phenomenon and the concept 
is not 
well defined for finite size systems.
Nevertheless, it has been demonstrated that clusters of pure helium
as small as 64 \he{} atoms exhibits clear 
``manifestations of superfluid behavior''~\cite{sindzingre_path-integral_1989}.
Moreover, the study of rotational constants of various molecules as
a function of the number $n$ of helium atoms has revealed the onset of
superfluid behavior for doped clusters composed of less than~10 \he{} 
atoms, leading to many studies on the concept of 
``microscopic molecular superfluid response''~\cite{tang_quantum_2002,tang_bridging_2004,mckellar_spectroscopic_2006,%
zeng_microscopic_2014}.
It is important to note that the area estimator,
which has been devised for sufficiently large clusters,
will necessarily 
overestimate the superfluid fraction for small clusters, since 
also
non-exchanging paths span a finite area.
In particular, in the limit of only one helium atom 
Eq.~(\ref{eq:fs-area}) leads to an unphysical finite superfluid 
fraction and for small clusters it can result in $f_{\rm s}$ values that are
systematically too large and even 
greater than unity.
The estimator is formally only valid in the thermodynamic limit 
for which the long exchange paths completely 
dominate the contributions due to the smaller non-exchanging paths.
Thus, Eq.~(\ref{eq:fs-area}) should
be used with great caution for very small clusters.
In order to circumvent this problem,
a rescaled estimator, that aims at removing the contribution 
of non exchanging paths from the area estimator,
has recently been introduced~\cite{zeng_probing_2013,%
zeng_microscopic_2014}.
This 
so-called ``exchange
estimator'' is defined as 
\begin{equation}
   f_{\rm s}^\alpha = \frac{I^\text{eff}_{\alpha,\text{MB}}-I^\text{eff}_{\alpha,\text{BE}}}{I^\text{eff}_{\alpha,\text{MB}}} 
              = 1 - \frac{I^\text{eff}_{\alpha,\text{BE}}}{I^\text{eff}_{\alpha,\text{MB}}}
   \label{eq:fs-exch}
\end{equation}
with $I^\text{eff}_{\text{MB}}$ the effective moment of inertia,
obtained from a path integral simulation in which the helium
follows Maxwell-Boltzmann statistics (i.e. without quantum exchange 
and thus neglecting the bosonic nature of \he{}) and 
$I^\text{eff}_{\text{BE}}$ is the effective moment of inertia 
obtained from a bosonic treatment of the helium.
This estimator is able to isolate the contribution to the area 
coming only from the exchanging paths and has thus been dubbed the 
``exchange''~(X) estimator.

\begin{figure}
   \centering
   \includegraphics[width=\textwidth]{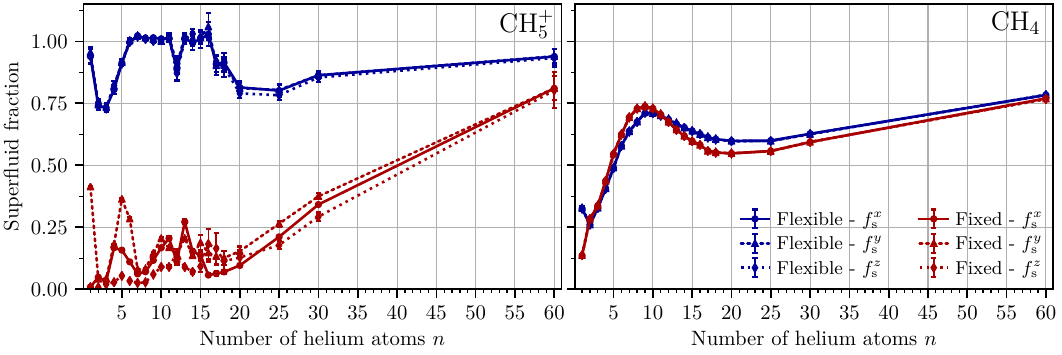}
   \caption{Superfluid fraction along the $x,y$ and $z$ direction
            at $T=0.5$\,K as a function of the number of helium 
            atoms around \pmeth{} (left) and \meth{} (right) 
            computed using the area estimator (Eq.~(\ref{eq:fs-area})). 
            The red lines corresponds to the results obtained 
            with a molecule completely fixed in its minimum 
            energy configuration, while the blue lines show
            the results obtained with a fully flexible impurity.}
   \label{fig:sff-1}
\end{figure}

Fig.~\ref{fig:sff-1} shows the superfluid fraction obtained using 
the standard area estimator as a function of the number of helium 
atoms computed in the space fixed frame (or laboratory frame).
The first observation is the striking difference between the
superfluid fractions obtained around a fully flexible and 
a completely fixed \pmeth{}, discussed in detail in the main
text of the manuscript.
Neglecting the molecular degrees of freedom of \pmeth{} results
in a significant 
suppression
of the superfluid fraction.
As discussed in detail in the main manuscript, this effect 
directly comes from the high degree of localization of \he{}
in the first shell that hinders the creation of long exchange 
cycles and thus suppresses superfluidity in the first shell.
Upon increasing the size of the cluster, the difference between 
the flexible and fixed case is considerably reduced~--
but only since the relative 
impact 
of what happens in 
the first shell on the total superfluid fraction 
of the entire cluster
decreases 
for large $n$ values. 
The effect remains, however, present
in the first shell
as can be seen from the 
local superfluid density (Fig.~\ref{fig:sfd-length}
and Fig.~4 of the main text)
for 
\pmeth{}$\cdot$He$_{60}$.
In the case of \meth{}, this impact of the molecular degrees of 
freedom is not found and the values of $f_{\rm s}$ obtained with a 
flexible or fixed impurity are very similar.
It is interesting to note as well that around the fixed \pmeth{}
molecule the obtained superfluid fraction is not isotropic due to the 
pronounced 
asymmetry of the \pmeth{}$\ccdots$He potential,
thus yielding different values $f^\alpha_{\rm s}$ 
for the three different directions $\alpha$. 
This effect is not present in the flexible case due to molecular rotations
and is also not seen around the fixed \meth{} molecule due to the high symmetry 
of the \meth{}$\ccdots$He potential.

For very small numbers of helium atoms, significant superfluid 
fractions are obtained around the flexible \pmeth{} and, 
in particular, a finite superfluid fraction is observed 
with a single
helium 
atom, which is a clear illustration of the failure of the area 
estimator for very small clusters
as is well-known from the literature. 
Upon increasing the number of helium atoms $n$ for \pmeth{}, the superfluid 
fraction quickly reaches unity for $6<n<16$ except for the special
case of $12$ \he{} atoms, discussed in detail in the following.
\begin{figure}
   \centering
   \includegraphics[width=0.6\textwidth]{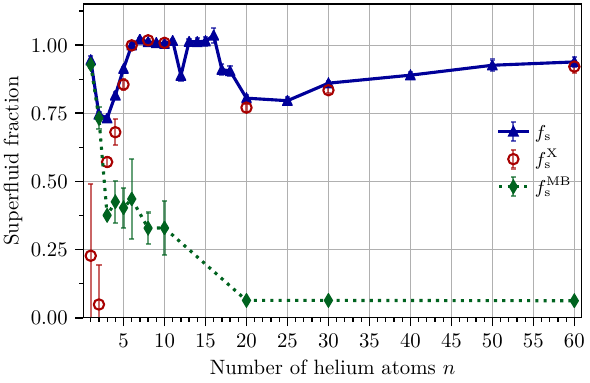}
   \caption{Superfluid fraction $f_\text{s}$ computed using 
            the area estimator (Eq.~(\ref{eq:fs-area})) and
            $f_\text{s}^\text{X}$ computed using the exchange 
            estimator (Eq.~(\ref{eq:fs-exch})) as a function 
            of the number of helium of atoms around flexible
            \pmeth{} at $T=0.5$\,K. 
            The superfluid fraction is computed as an 
            average over the three directions since $f_{\rm s}$ 
            is isotropic here. Additionally the superfluid 
            fraction 
formally
computed using the area estimator 
applied to non-bosonic Maxwell-Boltzmann simulations, 
            $f_\text{s}^\text{MB}$, is included as well to give an 
            estimation of the contribution of the non-exchanging 
            paths
to $f_\text{s}$ obtained from the area estimator.}
   \label{fig:sff-2}
\end{figure}
For $n>16$, the 
total superfluid fraction in the entire cluster
drops due to the creation 
of the second helium solvation shell, which features a smaller density that 
prevents the creation of long exchange paths.
Upon building this second solvation shell, the associated density 
increases 
again for $n \gg 16$
and so does that the superfluid fraction reaching around~0.9 
for the largest cluster of 60 \he{} atoms.
In the case of \meth{}, the superfluid fraction reaches a first maximum 
for $n=9$, which also corresponds to the beginning of the building of 
a second solvation shell (see Fig.~\ref{fig:shell-filling}), in which 
the density and thus the superfluid fraction is lower.
Upon building this second solvation shell, the superfluid fraction 
increases again to reach a value of around~0.75 for the largest cluster 
of 60 \he{} atoms.

In order to estimate the validity of the superfluid fractions computed here
using the area estimator, we have performed additional simulations without 
bosonic exchange for a few selected numbers of helium atoms around
flexible protonated methane in order to compute the superfluid fraction
using the exchange estimator~\cite{zeng_probing_2013,zeng_microscopic_2014}
of Eq.~(\ref{eq:fs-exch}).
The comparison between the superfluid fraction obtained using the two estimators
is presented in Fig.~\ref{fig:sff-2}.
It is clear that except for very small numbers of helium atoms, 
the area estimator provides a very good estimation of the 
superfluid fraction, since both estimators provide almost the 
same value for $n>5$ \he{} atoms
(thus the blue triangles and red circles are close to each other).
Only in the small cluster limit 
(in particular for 
$\le 4$), the exchange estimator
(red circles) 
yields substantially smaller 
$f_\text{s}$
values and eventually approaches the expected
limit, 
namely zero,
within error bars.

\begin{figure}[h]
   \centering
   \includegraphics[width=0.93\textwidth]{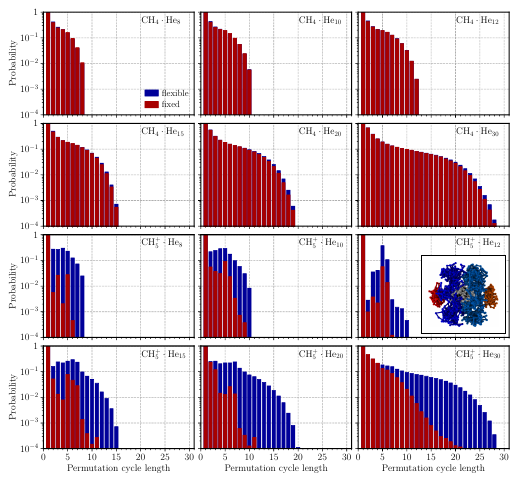}
   \caption{Probability of finding at least one exchange path of
            specific length for \meth{}$\cdot$\he$_n$ (first two rows)
            and \pmeth{}$\cdot$\he{}$_n$ (second two rows) at $T=0.5$\,K
for different clusters sizes $n$ as indicated. 
Data for the flexible and fixed impurities
are shown in blue and red, respectively;
note that blue is superimposed by red where not visible.
            The inset of \pmeth{}$\cdot$\he{}$_{12}$ depicts a representative
            snapshot from the simulation illustrating a typical arrangement 
            of the helium atoms for this very special case, see text. The helium beads 
            represented with the same color belongs to the same exchange cycle.}
   \label{fig:perm}
\end{figure}
It is well-known that superfluidity is directly connected to 
the
presence
of long exchange cycles~\cite{ceperley_path_1995,%
krauth_quantum_1996,%
kwon_atomic-scale_1999}, 
i.e. of path lengths comparable to the system size,
and the statistics of exchange path length, as shown in 
Fig.~\ref{fig:perm}, can thus provide
complementary
insight into the origin of the superfluid response.
The exchange path statistics clearly confirm that, in the case of \meth{},
the molecular degrees of freedom have no significant impact on bosonic 
exchange and thus on the superfluid behavior of helium.
In the case of \pmeth{} however, there is a clear coupling between the 
molecular motion and bosonic exchange and, in particular, neglecting the 
molecular degrees of freedom 
by fixing \pmeth{} in space
tends to 
significantly
reduce the probability of building 
long exchange cycles, which reduces the superfluid response.

A special case appears for \pmeth{}$\cdot$He$_{12}$, for which a 
sudden drop of the superfluid fraction is also noticed in 
Fig.~\ref{fig:sff-1}.
This effect is related to a particularly stable configuration of the 
helium atoms, forming two parallel rings of five atoms around \pmeth{} 
with the last two atoms being located at two opposite ends of the 
molecule, as shown in the inset of \pmeth{}$\cdot$\he$_{12}$ in 
Fig.~\ref{fig:perm}.
This configuration in Cartesian space leads to a special distribution 
in permutation space as well, with the exchange path length of five 
being overrepresented, as such rings are mostly found in exchange cycles
involving all five atoms.
Similar configurations have been previously observed for instance in 
para-H$_2$ clusters doped with a water molecule~\cite{%
zeng_simulating_2013}, thus indicating that this specific topology of 
the exchange path might be a universal feature of doped bosonic 
clusters featuring a strong molecule$\ccdots$solvent interaction. 
\subsection{Local Superfluid Density}
\begin{figure}
   \centering
   \includegraphics[width=\textwidth]{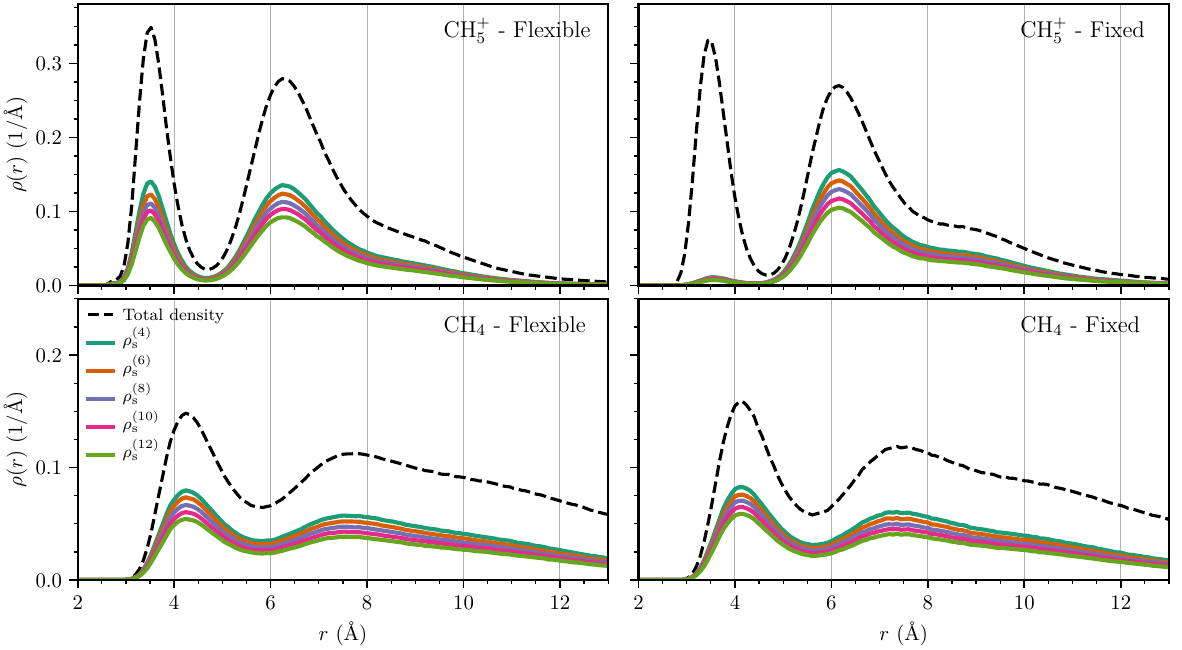}
   \caption{%
Local superfluid density $\rho_{\rm s}^{(l)}$ 
for \pmeth{}$\cdot$He$_{60}$ 
            (first row) \meth{}$\cdot$He$_{60}$ (second row) 
            computed using the ``local exchange path estimator'' of 
            Eq.~(\ref{eq:sfd-length}) at $T=0.5$\,K using 
            different values of the cutoff length $l$
where $r$ is the distance between the carbon nucleus with respect to helium, 
see text.
Left and right columns are for flexible and fixed molecules, respectively. 
}
   \label{fig:sfd-length}
\end{figure}

The well-known connection between the presence of long exchange paths
and superfluidity~\cite{ceperley_path_1995,krauth_quantum_1996,%
kwon_atomic-scale_1999} 
can actually be used as a proxy to obtain local information on the superfluid
behavior of the helium solvent.
In particular,
an estimator of the local superfluid density $\rho_{\rm s}(r)$
was 
introduced~\cite{kwon_atomic-scale_1999}
based on the distributions of helium atoms 
involved in an exchange path longer than a certain cutoff length $l$,
\begin{equation}
   \rho_{\rm s}^{(l)}(\vec{r}) = \sum_{p > l}^{N_\text{He}} \rho_p(\vec{r})
\enspace , 
   \label{eq:sfd-length}
\end{equation}
where $\rho_p(\vec{r})$ is the local density of helium 
involved in  an exchange cycle of length~$p$
at point $\vec{r}$ in space. 
Figure~\ref{fig:sfd-length} displays the local superfluid density obtained using 
this 
``local exchange path estimator''
for different $l$~parameters 
where $r$ is the distance between the carbon nucleus in \pmeth{} and \meth{}
with respect to helium. 
In particular, the impact of the choice of the cutoff to distinguish between 
``short'' and ``long'' exchange path is investigated in the following.
While the obtained superfluid densities clearly depend on the value 
of the cutoff parameter~$l$, their shapes remain unchanged for different values
of $l$ and Eq.~(\ref{eq:sfd-length}) thus provides a relatively robust
estimator of the superfluid density. 
The results presented in the main manuscript are obtained using a cutoff of $l=6$.
However, the estimator tends to systematically underestimate the
superfluid density, as expected based on its definition.
Short paths,
and in particular non-exchanging ones, strongly
contribute to the normal density in this estimator,
even though such paths are also present in a
fully superfluid case.
Nevertheless,
this estimator of the local superfluid density 
still grants access to a well-defined
locally resolved estimation of the superfluid response
as we will demonstrate in what follows by comparing the results 
provided by this estimator (as discussed in the main manuscript)
with those using another local estimator. 

An alternative local estimator based on the area of the Feynman paths 
in a similar way as the (global) area estimator of the superfluid fraction 
has also been developed~\cite{draeger_superfluidity_2003,kwon_local_2006}.
In particular, a ``local area estimator'' of the 
superfluid density 
has been introduced,
which leads to the correct value of the effective
moment of inertia~\cite{kwon_local_2006}
\begin{eqnarray}
   I_{\alpha}^\text{eff} & = \int \left( \rho(\vec{r}) - \rho^\alpha_s(\vec{r}) \right) r^2_{\perp} \text{d}^3\vec{r}\\
   I_{\alpha}^\text{eff} & = I_{\alpha}^\text{cl} - \int \rho^\alpha_s(\vec{r}) r^2_{\perp} \text{d}^3\vec{r} .
\end{eqnarray}
The last term in the right hand side clearly represents the quantum 
reduction of the effective moment inertia due to superfluidity and 
is given by 
\begin{equation}
   \int \rho^\alpha_s(\vec{r}) r^2_{\perp} \text{d}^3\vec{r} = \frac{4m_\text{He}^2}{\beta\hbar^2} \langle A_\alpha^2 \rangle,
\end{equation}
from which the following superfluid density estimator 
can be obtained~\cite{kwon_local_2006}
\begin{equation}
   \rho^\alpha_s(\vec{r}) = \frac{4m_\text{He}^2 \langle A_\alpha A_\alpha(\vec{r})\rangle}{\beta\hbar^2 r^2_{\perp}},
\end{equation}
with $A_\alpha(\vec{r})$ the local contribution to the 
area defined as
\begin{equation}
   A_\alpha(\vec{r}) = \frac{1}{2}\sum_{i=1}^{N_\text{He}}\sum_{s=1}^P \left(\vec{r}_{i,s} \times \vec{r}_{i,s+1}\right)_{\alpha}
                       \delta(\vec{r}_{i,s}-\vec{r}). \label{eq:area}
\end{equation}
A local superfluid fraction can then be defined as well, 
\begin{equation}
   \frac{\rho^\alpha_s(\vec{r})}{\rho(\vec{r})} = 
   \frac{4m_\text{He}^2 \langle A_\alpha A_\alpha(\vec{r})\rangle}{\beta\hbar^2 I^\text{cl}(\vec{r})}
, 
   \label{eq:fs-local-area}
\end{equation}
which can be used to compute the values of superfluid 
fraction associated with different regions, such as
the first and second shell, by locally averaging the
contribution to the local exchange area and moment of 
inertia~\cite{kwon_local_2006}.
This can be used furthermore to define a radial superfluid 
fraction or equivalently a radial superfluid density, 
\begin{equation}
   \rho^\alpha_\text{s}(r) = \frac{4m_\text{He}^2 \langle A_\alpha A_\alpha(r)\rangle}{\beta\hbar^2 I^\text{cl}(r)} \rho(r) 
   \label{eq:sfd-area}
\end{equation}
as shown in the bottom panel of Fig.~4 of the main text.
Note that the distance $r$ corresponds here to the distance 
of helium with respect 
to the center of 
mass of the molecule computed using all the beads
(rather than to the C~atom of the respective molecule as used in Fig.~\ref{fig:sfd-length}). 
The differences between the total helium density
in Fig.~4 of the main text are
simply
due to this change of 
the reference point for computing distances,
noting that the center of mass 
distance is a more natural choice for the local area estimator.
Indeed, the vectorial product in Eq.~(\ref{eq:area}) requires
a common origin for the two position vectors, which should thus not 
depend on the bead index and we chose here to use the molecular center 
of mass computed using all the beads.

As mentioned in the main text, both estimators
provide the same overall picture for both impurities, 
in particular the distance modulations of the superfluid
response within the first and second helium shells 
when comparing the flexible to the fixed impurity scenario. 
Yet, despite
providing 
such 
qualitative agreement, the
local area estimator leads to superfluid densities that 
are typically quantitatively different in numbers
from the ones obtained using the ``local exchange path estimator'' (see 
Fig.~\ref{fig:sfd-length}).
In particular, with this estimator, we find that the global superfluid 
fraction associated with the \meth{}$\cdot$He$_{60}$ complex is 0.86 which is 
larger in number
than the value of 0.78 obtained using the associated 
global
superfluid fraction estimator of Eq.~(\ref{eq:fs-area}).
In the case of the \pmeth{}$\cdot$He$_{60}$ complex, the obtained 
superfluid density indicates that the cluster is essentially fully 
superfluid around flexible \pmeth{} with an associated global superfluid
fraction of 0.96, which is 
indeed very 
similar to the value of 0.94 that we obtained 
using the global estimator of Eq.~(\ref{eq:fs-area}).
In particular, the first solvation shell is fully superfluid.
In the case of the fixed protonated methane, the obtained superfluid 
density indicates, as expected, a much lower superfluid fraction estimated 
to be around 0.68 which is even smaller than the value of 0.78 obtained 
with the global estimator.
The superfluid density in the first shell is particularly impacted
and the associated superfluid fraction dropped to 0.28 as a result 
of the strong localization of helium as  discussed in the main manuscript.

In summary, both estimators fully consistently disclose that 
sufficiently large 
\pmeth{}$\cdot$He$_n$ 
complexes feature a strong superfluid response of the first 
(frozen) solvation 
shell
in the fully flexible case, 
and thus manifestations of supersolid behavior,
which is
lost if the 
large-amplitude 
ro-vibrational
degrees of freedom
of the molecule
are neglected
by fixing all intramolecular molecular degrees of freedom,
as discussed in detail in the main manuscript.
Recall that this phenomenon, which can only be
observed with sufficiently many helium atoms
to fully solvate the molecular impurity,
thus filling the first solvation shell,
cannot appear in the microsolvation limit and
is 
thus 
strikingly different
from what has been discovered earlier~\cite{uhl_quantum_2019} 
for 
very small
\pmeth{}$\cdot$He$_n$
up to $n=4$. 
\subsection{Microsolvation Limit}
The microsolvation of \pmeth{} up to a maximum number of four helium atoms has 
been studied some time back~\cite{uhl_helium_2018,uhl_quantum_2019}
which we discuss in the following in relation to our current work 
that focuses on novel effects of this impurity seen in the realm of large \he{} clusters. 
Previously, an intricate coupling 
between the molecular motion of the impurity and bosonic exchange of helium 
has been discovered for such tiny \pmeth{}$\cdot$\he$_n$ clusters with $n$ up to 
four helium atoms~\cite{uhl_quantum_2019}.
Such small system sizes $n$ are of particular interest to
\he{}-based tagging spectroscopy~\cite{roithova_helium_2016,Topfer2018/10.1103/PhysRevLett.121.143001,Asvany2019/10.1021/acs.jpclett.9b01911}. 
Given the lack of a well-defined, unique estimator to compute 
the superfluid fraction in the limit of a small number
of \he{} atoms (as discussed here in Sec.~\ref{sec:fs}), 
which is particularly relevant for \pmeth{}$\cdot$\he$_n$ 
with $n\leq 4$, the previous analysis~\cite{uhl_quantum_2019} 
has been performed entirely at the qualitative level.
In particular, the exchange path statistics of those paths that include all available $n$ \he{} atoms,
i.e. as few as $n=2$, 3 and 4 in this case, has been studied; 
this approach closely follows pioneering work~\cite{krauth_quantum_1996,kwon_atomic-scale_1999}
on using ``sufficiently long'' exchange cycles as a proxy to detect superfluid behavior.

Here,  in stark contrast, we are interested 
in the opposite limit to microsolvation, namely
completely filling the first solvation shell and 
going beyond by studying large clusters with up to $n=60$ \he{} atoms
that solvate the CH$_5^+$ molecule. 
We demonstrate based on Fig.~\ref{fig:sff-2} that
the area and exchange estimators reviewed in Sec.~\ref{sec:fs} provide very similar
superfluid fractions only for $n\geq 6$ which
remains so all the way up to the largest cluster $n=60$.
We therefore refrain here from discussing the superfluid fraction
for \pmeth{}$\cdot$\he$_n$ clusters with $n<6$
given the strongly different $f_\text{s}$ results and,
thus, ambiguity in the microsolvation limit. 
Moreover, we study here only either the fully flexible
or the frozen CH$_5^+$ core within the  \pmeth{}$\cdot$\he$_n$ clusters,
whereas major insights on the impact of large-amplitude motion
on bosonic exchange within microsolvated clusters with $n\leq 4$ 
were enabled in the previous study~\cite{uhl_quantum_2019}
by performing so-called ``e--C$_{\rm s}$ restained simulations'' 
to generate data on a useful reference system 
(namely the one that allows only for small-amplitude motion
while not being fixed in space). 
This aspect is not at all in the focus of the present investigation on 
\pmeth{}$\cdot$\he$_n$ clusters that 
is devoted to what happens in the limit of large clusters,
in particular beyond having filled the first solvation shell.
For this purpose, we introduce \meth{}$\cdot$\he$_n$ clusters to provide the
proper reference system to assess effects due to large-amplitude
motion (which only operate in \pmeth ) since \meth{} is a standard
quasi-rigid molecule subject to
small-amplitude motion that is well described by
quasi-harmonic deviations from a unique equilibrium structure.

Apart from that, we achieved major improvements
of the accuracy of the interactions, 
i.e. the \pmeth{} potential energy surface 
and  the \pmeth{}$\ccdots$\he{} interaction potential
that we can now use compared to previously~\cite{uhl_quantum_2019}. 
In that previous work, a simple force field description of the 
\pmeth{} potential energy surface has been used, 
the so-called POSflex model~\cite{uhl_adding_2014}, 
where all five C--H bond distances have the same
equilibrium length, thus all protons move on a sphere
centered around the carbon site. 
This force field is computationally highly efficient,
but only allows for a merely qualitatively satisfactory 
description of the properties of \pmeth{} 
as demonstrated previously~\cite{uhl_adding_2014}, 
since POSflex does not take into account the characteristically
different lengths~\cite{marx_structural_1995} of the three-center 
versus two-center C--H bonds within CH$_5^+$. 
Secondly, 
the potential constructed previously
to describe the \pmeth{}$\ccdots$\he{} interactions
is based on a complicated, traditional force field-like representation
fitted against coupled cluster interaction
energies~\cite{kuchenbecker_constructing_2017}.
Unfortunately, this potential is only able to provide 
a satisfactory description of 
very small \pmeth{}$\cdot$\he$_n$ clusters up to only four 
\he{} atoms~\cite{kuchenbecker_constructing_2017}, 
whereas it has been found to fail even qualitatively for $n>4$.
Thus, the present study would not have been possible using
this existing interaction potential.

In the current work, in stark contrast to our previous investigation~\cite{uhl_quantum_2019}, 
we greatly improved the quality
of all interactions, both intra- and intermolecular, 
and pushed them to the converged
coupled cluster level, CCSD(T), enabled by using 
our high-dimensional neural network potential approach 
for finite systems~\cite{schran_high-dimensional_2017,schran_automated_2020} 
at what is often called ``chemical accuracy''. 
This major development is comprehensively presented 
in a self-contained manner in Sec.~\ref{sec:nnps}
where we describe in detail the generation and validation of the
CH$_5^+$ and CH$_4$ potential energy surfaces
as well as the \pmeth{}$\ccdots$\he{} 
and \meth{}$\ccdots$\he{} interaction potentials
that we introduced and used in the present investigation
for the first time.

Another significant improvement of the present
study 
versus
our previous work~\cite{uhl_quantum_2019} 
is related to the colored-noise thermostatting scheme
within the path integral approach to effectively enhance 
the convergence of the Trotter decomposition. 
The currently used PIQTB technique~\cite{brieuc_quantum_2016},
carefully validated for use at very low temperatures~\cite{schran_zundel_2018},
allows us to reach a temperature of 0.5~K in a 
path integral molecular dynamics setup, whereas
previously, using the PIGLET technique~\cite{Ceriotti2012/10.1103/PhysRevLett.109.100604,Uhl2016/10.1063/1.4959602},
we were able to reach only 1.25~K, where bosonic 
exchange effects are of course much less pronounced. 
Overall, our present quantum simulation approach significantly transcends,
both in accuracy and convergence, the one used a few years back~\cite{uhl_quantum_2019}
to study microsolvated \pmeth{}$\cdot$\he$_n$ clusters with $n\leq 4$
in many ways.
Only these significant improvements grant access to accurately
simulating large \pmeth{}$\cdot$\he$_n$  clusters~-- which
provides the basis to discover a novel effect, namely
\textit{manifestations of supersolid behavior}.
We refer the interested reader to a recent review where all these
major improvement of our neural network-based 
bosonic path integral simulation methodology,
which enables converged quantum simulations of complex
molecular systems such as the present one at very low temperatures,
are explained and validated in detail~\cite{brieuc_converged_2020}.

\clearpage